\documentclass[8.5pt,twoside,twocolumn]{article}
\usepackage[super,sort&compress,comma]{natbib} 
\usepackage{times,mathptmx}
\usepackage{sectsty}
\usepackage{balance} 

\usepackage{graphicx} %eps figures can be used instead
\usepackage{lastpage}
\usepackage[format=plain,justification=raggedright,singlelinecheck=false,font=small,labelfont=bf,labelsep=space]{caption} 
\usepackage{fancyhdr}
\usepackage{threeparttable}
\usepackage{multirow}
\usepackage{enumitem}
\usepackage{url}
\usepackage[tight,footnotesize]{subfigure}

\begin{document}

\thispagestyle{plain}
\fancypagestyle{plain}{
\fancyhead[L]{\includegraphics[height=8pt]{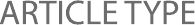}}
\fancyhead[C]{\hspace{-1cm}\includegraphics[height=20pt]{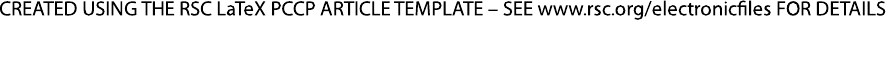}}
\fancyhead[R]{\includegraphics[height=10pt]{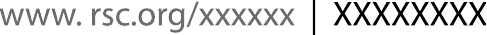}\vspace{-0.2cm}}
\renewcommand{\headrulewidth}{1pt}}
\renewcommand{\thefootnote}{\fnsymbol{footnote}}
\renewcommand\footnoterule{\vspace*{1pt}% 
\hrule width 3.4in height 0.4pt \vspace*{5pt}} 
\setcounter{secnumdepth}{5}

\makeatletter 
\def\subsubsection{\@startsection{subsubsection}{3}{10pt}{-1.25ex plus -1ex minus -.1ex}{0ex plus 0ex}{\normalsize\bf}} 
\def\paragraph{\@startsection{paragraph}{4}{10pt}{-1.25ex plus -1ex minus -.1ex}{0ex plus 0ex}{\normalsize\textit}} 
\renewcommand\@biblabel[1]{#1}            
\renewcommand\@makefntext[1]% 
{\noindent\makebox[0pt][r]{\@thefnmark\,}#1}
\makeatother 
\renewcommand{\figurename}{\small{Fig.}~}
\sectionfont{\large}
\subsectionfont{\normalsize} 

\fancyfoot{}
\fancyfoot[LO,RE]{\vspace{-7pt}\includegraphics[height=9pt]{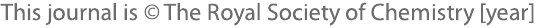}}
\fancyfoot[CO]{\vspace{-7.2pt}\hspace{12.2cm}\includegraphics{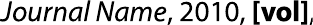}}
\fancyfoot[CE]{\vspace{-7.5pt}\hspace{-13.5cm}\includegraphics{RF}}
\fancyfoot[RO]{\footnotesize{\sffamily{1--\pageref{LastPage} ~\textbar  \hspace{2pt}\thepage}}}
\fancyfoot[LE]{\footnotesize{\sffamily{\thepage~\textbar\hspace{3.45cm} 1--\pageref{LastPage}}}}
\fancyhead{}
\renewcommand{\headrulewidth}{1pt} 
\renewcommand{\footrulewidth}{1pt}
\setlength{\arrayrulewidth}{1pt}
\setlength{\columnsep}{6.5mm}
\setlength\bibsep{1pt}

\twocolumn[
  \begin{@twocolumnfalse}
\noindent\LARGE{\textbf{Metals Production Requirements for Rapid Photovoltaics Deployment$^\dag$}}
\vspace{0.6cm}

\noindent\large{\textbf{Goksin Kavlak, \textit{$^{a}$} James McNerney, \textit{$^{a}$} 
Robert L. Jaffe \textit{$^{b}$} and Jessika E. Trancik \textit{$^{a,c,\ast}$}}}\vspace{0.5cm}
%Please note that \ast indicates the corresponding author(s) but no footnote text is required. 

%\noindent\textit{\small{\textbf{Received Xth XXXXXXXXXX 20XX, Accepted Xth XXXXXXXXX 20XX\newline
%First published on the web Xth XXXXXXXXXX 200X}}}

%\noindent \textbf{\small{DOI: 10.1039/b000000x}}
%\vspace{0.6cm}
%Please do not change this text.

\noindent \normalsize{If global photovoltaics (PV) deployment grows rapidly, the required input materials need to be supplied at an increasing rate. In this paper, we quantify the effect of PV deployment levels on the scale of metals production. For example, we find that if cadmium telluride \{copper indium gallium diselenide\} PV accounts for more than 3\% \{10\%\} of electricity generation by 2030, the required growth rates for the production of indium and tellurium would exceed historically-observed production growth rates for a large set of metals. %Other thin-film PV metals, such as gallium and selenium, would require growth rates that are on the higher end of the past growth rates. 
In contrast, even if crystalline silicon PV supplies all electricity in 2030, the required silicon production growth rate would fall within the historical range.} %Besides the strikingly high projected growth rates, this analysis shows that the required levels of annual production for thin-film PV metals exceed their potentially available and recoverable levels.}
More generally, this paper highlights possible constraints to the rate of scaling up metals production for some PV technologies, and outlines an approach to assessing projected metals growth requirements against an ensemble of past growth rates from across the metals production sector. The framework developed in this paper may be useful for evaluating the scalability of a wide range of materials and devices, to inform technology development in the laboratory, as well as public and private research investment.

\vspace{0.5cm}
 \end{@twocolumnfalse}
  ]

%Footnotes
\footnotetext{\dag~Electronic Supplementary Information (ESI) available.}

%Please use \dag to cite the ESI in the main text of the article.
%If you article does not have ESI please remove the the \dag symbol from the title and the above footnotetext.

\footnotetext{\textit{$^{a}$~Engineering Systems Division, Massachusetts Institute of Technology, Cambridge, MA, 02139, USA.}}
\footnotetext{\textit{$^{b}$~Center for Theoretical Physics and Department of Physics, Massachusetts Institute of Technology, Cambridge, MA, 02139, USA. }}
\footnotetext{\textit{$^{c}$~Santa Fe Institute, Santa Fe, NM, 87501, USA. }}
\footnotetext{\textit{$^{\ast}$~trancik@mit.edu }}

%additional addresses can be cited as above using the lower-case letters, c, d, e... If all authors are from the same address, no letter is required

%\footnotetext{\ddag~Additional footnotes to the title and authors can be included \emph{e.g.}\ `Present address:' or `These authors contributed equally to this work' as above using the symbols: \ddag, \textsection, and \P. Please place the appropriate symbol next to the author's name and include a \texttt{\textbackslash footnotetext} entry in the the correct place in the list.}

\section{Introduction} \label{sec:introduction}

Photovoltaics (PV) is a low-carbon technology that has the potential to reduce greenhouse gas emissions if deployed at large scale. \cite{Pacala2004, Riahi2007, Trancik2013} As of 2012, PV provides only 0.4\% of the world's electricity. \cite{IEA2014} Its deployment is growing rapidly, however, at an average rate of 30\% per year, \cite{Trancik2014} as the technology steadily improves and costs decline. \cite{Trancik2006, Nagy12, Farmer07b, Bettencourt2013, Zheng2014}

The future growth of PV has been estimated in various energy scenarios, based on projections of energy demand and the cost and performance of technologies in the future. Various international organizations, \cite{IEA2012, IEA2010} environmental agencies and industry associations, \cite{GWEC2012, EPIA2011} energy companies and other corporations \cite{Shell2008, BNEF2013} and academic institutions and researchers \cite{GEA2012, Jacobson2011} have contributed to this literature. Another group of studies focuses on resource constraints and the potential for future PV deployment. For example, various researchers have analyzed the material constraints on PV deployment that are imposed by annual metal production levels or reserves \cite{Andersson2000, Wadia2009, Kleijn2011, Tao2014} and have discussed the potential for increasing PV deployment by reducing the material intensity of PV technologies. \cite{Feltrin2008, Woodhouse2013a, Tao2014, Zuser2011} 

While these studies address the production scale of metals eventually needed, they do not directly address the time frame over which scaling up should be achieved. In this paper, we ask whether metals production can be scaled up at a pace that matches the rapidly increasing PV deployment levels put forward in aggressive low-carbon energy scenarios. Based on the projected PV deployment levels in 2030, we estimate the growth rates required for metals production to satisfy the metal demand by the PV sector. We present a new perspective on the metal requirements of PV deployment by comparing the required growth rates with the growth rates observed in the past by a large set of metals (the full set of metals for which yearly production data is available for all years in the period 1972 to 2012). (See Section 1 of Supplementary Information for details.\dag) %Goksin, please check this statement. GK: Correct. I added SI. 

We include in our analysis the absorber layer materials of three PV technologies manufactured and sold today: crystalline silicon (c-Si) technology (roughly 90\% of annual PV production today) \cite{Fraunhofer2014} and two thin-film PV technologies, cadmium telluride (CdTe) and copper indium gallium diselenide (CIGS) (with roughly 5\% and 2\% of annual PV production today),  \cite{Fraunhofer2014} building on our earlier, preliminary results. \cite{Kavlak2014} Whereas c-Si is based on an abundant metal, silicon, CdTe and CIGS utilize metals that have low crustal abundance and are obtained as byproducts of other metals' production. %Goksin, please check market shares. GK: Shares refer to the share in annual global production in 2013. Shares are: 91% c-Si, 5% CdTe, 2% CIGS, and 2% amorphous-Si. (Fraunhofer 2014 PV report)

In this paper we aim to provide a thorough analysis of the required growth rates for silicon in c-Si, tellurium in CdTe, and indium, gallium and selenium in CIGS to meet a range of projected PV growth scenarios. To complement this analysis of past and projected metals growth rates, we also compare the projected levels of metals production to their estimated scalability potential based on metals reserves. The approach developed in this paper may also be useful for studying the scalability potential of other technologies as well, in light of the production growth requirements of raw materials.

%In this study, we extend our analysis to (1) include cadmium in order to complete the analysis of the metals requirements of CdTe, (2) include further study of changes in historical growth rates of metals production over time, and (3) compare the projected levels of metals production to their estimated scalability potential based on metals reserves.

%%%%%%%%%%%%%%%%%%%%%%%%%%%%%%%%%%%%%%%%%%%%%%%%%%%%%%%%%%%%%%
\section{Methods}\label{sec:Methods} 

We estimate the growth rates required for metals production to meet the metal demand associated with projected global PV deployment levels in 2030. These projected levels are based on a number of published energy scenarios ranging from low to high PV deployment (table \ref{Table1}). We note that providing a high proportion of the total electricity through PV would require energy storage technologies that would also entail material requirements. \cite{Barnhart2013} This paper concentrates on the materials used in PV technologies but could be extended to analyze energy storage technologies.

The analysis begins with estimating the required annual production in 2030 for each PV metal of interest. We then calculate the annual growth rate needed for the metals production to reach the required level in 2030. To estimate the required metal production in 2030, we consider the projected demand for the metal by both the PV sector and non-PV end-use sectors of the metal,

\begin{equation}
P_{\beta}= X_{\alpha}I_{\alpha\beta} + N_{\beta}(1+n_{\beta})^{18}
\label{eq:prod2030}
\end{equation}
where
\begin{description} [leftmargin=3em,style=nextline, noitemsep]
\item[$P_{\beta}$] required production for metal $\beta$ in 2030 [metric tons/year (t/y)]
\item[$X_{\alpha}$] deployment for PV technology $\alpha$ during 2030 [GW/y]
\item[$I_{\alpha\beta}$] intensity of metal $\beta$ for PV technology $\alpha$ [t/GW]
\item[$N_{\beta}$] metal $\beta$ used by non-PV end-uses in 2012 [t/y]
\item[$n_{\beta}$] annual growth rate in non-PV end-uses of metal $\beta$ [unitless]
\end{description}
\vspace*{9pt}

The projected demand for a metal $\beta$ by a PV technology $\alpha$ in 2030 is determined by the projected annual deployment level of the PV technology in 2030, $X_{\alpha}$, and the anticipated material intensity of metal $\beta$ in 2030, $I_{\alpha\beta}$. We calculate the annual PV deployment in 2030, $X_{\alpha}$, by using the cumulative installed PV capacity for 2030 projected by the energy scenarios (table \ref{Table1}) and assuming a constant percent annual growth from 2012 to 2030. The material intensity, $I_{\alpha\beta}$ (in g/W or t/GW), for a metal in a PV module is given by

\begin{equation}
I_{\alpha\beta}= \frac{t_{\alpha}\rho_{\alpha}w_{\alpha\beta}}{\sigma\eta_{\alpha}U_{\alpha\beta} y_{\alpha}}
\label{eq:matInt}
\end{equation}
where

\begin{description} [leftmargin=3em,style=nextline, noitemsep]
\item[$t_{\alpha}$] thickness of absorber layer for PV technology $\alpha$ [$\mu$m]
\item[$\rho_{\alpha}$] density of layer for PV technology $\alpha$ [g/cm$^{3}$]
\item[$w_{\alpha\beta}$] mass fraction of metal $\beta$ within the layer for PV technology $\alpha$ [unitless]
\item[$\eta_{\alpha}$] module efficiency for PV technology $\alpha$ [unitless]
\item[$\sigma$] solar constant [1000 W/m$^{2}$]
\item[$U_{\alpha\beta}$] utilization fraction of metal $\beta$ in manufacturing PV technology $\alpha$ [unitless]
\item[$y_{\alpha}$] yield in cell and module manufacturing for PV technology $\alpha$ [unitless]
\end{description}
\vspace*{9pt}

%%%%%%%%%%%%%%%%%%%%%%%%%%%%%%%
\begin{table}%[e]
%\captionsetup{font=normal}
\centering
\begin{threeparttable}
\renewcommand{\arraystretch}{1.3}
\caption{Cumulative Installed PV Capacity Projections for 2030.}
\label{Table1}

\begin{tabular}{@{}lcc}%{@{}l||c|c}
%\begin{tabular}{ c || c p{2cm}| c p{2cm} }
%\begin{tabular}{*{4}{p{3cm}}}
\hline
Energy Scenario & \shortstack{Cumulative\\installed PV\\ capacity (GW)} & \shortstack{Approximate \% of\\global electricity\\from PV} \\
\hline%\hline
IEA WEO\textsuperscript{\textit{a}} & 720 & 3 \\
Solar Generation 6\textsuperscript{\textit{b}} & 1850 & 8\\
GEA\textsuperscript{\textit{c}} & 3000 & 13\\
Shell\textsuperscript{\textit{d}} & 5500 & 24\\
\hline
\end{tabular}
\begin{tablenotes}
\item $^{\rm a}$ 450 scenario \cite{IEA2012}
\item $^{\rm b}$ Paradigm shift scenario \cite{EPIA2011}
\item $^{\rm c}$ GEA-supply, conventional transportation, full portfolio scenario \cite{GEA2012} 
\item $^{\rm d}$ Scramble scenario \cite{Shell2008}
\item Note: Installed capacity figures rounded to nearest ten GW. Approximate percentage of global electricity is calculated assuming 15\% capacity factor for PV \cite{EPIA2011} and a total global electricity generation of 30000 TWh in 2012. \cite{IEA2012}
%\item \gk{\textbf{Note: Maybe we can omit this table. The cumulative installed capacity values can be mentioned in the text, and annual installations in 2030 are already shown as vertical lines in the required rates figure in the Results section.}}
\end{tablenotes}
\end{threeparttable}
\end{table}

%%%%%%%%%%%%%%%%%%%%%%%%%%%%%%%
\begin{table*}\centering 
\begin{threeparttable}[e]
%\captionsetup{font=normal}
\renewcommand{\arraystretch}{1.3}
\caption{Parameters for Material Intensity and the Resulting Material Intensity, $I_{\alpha\beta}$, for Each Element.}
\label{Table2}
\begin{tabular}{llllllllll}
    %\toprule 
    \hline
Elements & Cases & $t_\alpha$ ($\mu$m) & $\eta_{\alpha}$ (\%) & $U_{\alpha\beta}$ (\%) & $y_\alpha$ (\%) & $\rho_\alpha$ (g/cm$^{3}$) & $w_{\alpha\beta}$ (\%) & $I_{\alpha\beta}$ (t/GW)\\ 
    \hline      						     
\multirow{3}{*}{In in CIGS} & high & 2  & 14  & 75  & 73  &\multirow{3}{*}{5.75}  & \multirow{3}{*}{22} & 28 \\
    						    & medium    & 1.2  & 15.7  & 80  & 90  &              &  				 & 13 \\ 
    						    & low  & 1.1  & 20  & 95 & 98   &                     & 				 & 7 \\ \hline  
    						     
\multirow{3}{*}{Ga in CIGS} & high & 2 & 14  & 75  & 73  &\multirow{3}{*}{5.75}  & \multirow{3}{*}{7}  & 9 \\
    						     & medium     & 1.2  & 15.7   & 80    & 90    &        & 				     & 4 \\ 
    						     & low  & 1.1  & 20  & 95    & 98    &             &  				   & 2 \\ \hline  
    						     		     
\multirow{3}{*}{Se in CIGS}  & high & 2  & 14 & 30 & 85  & \multirow{3}{*}{5.75}  & \multirow{3}{*}{50} & 161 \\
    						     & medium    & 1.2  & 15.7  & 60  & 90 &              &  				   & 41 \\ 
    						     & low  & 1.1  & 20  & 95  & 98  &                 &  				   & 17 \\ \hline  
    						     
\multirow{3}{*}{Te in CdTe} & high     & 2.5  & 11.7  & 50  & 85  &\multirow{3}{*}{5.85} & \multirow{3}{*}{53} & 156 \\ 
    						    & medium    & 2    & 14   & 70  & 90  &   			   &                        & 70 \\ 
    						    & low      & 1   & 18   & 95  & 97  &                       &                     & 19 \\ \hline   
   
\multirow{3}{*}{Cd in CdTe} & high & 2.5  & 11.7  & 50  & 85  &\multirow{3}{*}{5.85} & \multirow{3}{*}{47} & 138 \\ 
    						    & medium    & 2    & 14   & 70  & 90  &   			   &                     & 62 \\ 
    						    & low  & 1   & 18   & 95  & 97  &                       &                     & 17 \\ \hline     						    
\multirow{3}{*}{Si in c-Si}  & high & 180  & 14.8   & 45   & 95  &  \multirow{3}{*}{2.33} & \multirow{3}{*}{100} & 6629 \\ 
    						    & medium     & 120  & 18    & 55    & 98  &                  &                      & 2882 \\ 
    						    & low  & 50   & 20.5  & 90    & 99  &                        &                      & 638 \\ \hline    

\end{tabular}			       						    

\begin{tablenotes}
%\item References:

\item \textbf{In, Ga, Se:} $t_{\alpha}$ high \cite{Woodhouse2013a, Zuser2011}; $t_{\alpha}$ medium, $t_{\alpha}$ low \cite{Zuser2011}; $\eta_\alpha$ high \cite{Miasole}; $\eta_\alpha$ medium, $\rho_\alpha$, $w_{\alpha\beta}$ for In, $w_{\alpha\beta}$ for Ga \cite{Woodhouse2013a}; $\eta_\alpha$ low \cite{Marwede2014}; $U_{\alpha\beta}$ high for In, $U_{\alpha\beta}$  high for Ga \cite{Woodhouse2013a}, other $U_{\alpha\beta}$ values \cite{Marwede2014}; $w_{\alpha\beta}$ for Se \cite{Kamada2010}; $y_\alpha$ high for In, $y_\alpha$ high for Ga \cite{Woodhouse2013a}, other $y_\alpha$ values \cite{Marwede2014}.

\item \textbf{Te, Cd:} $t_{\alpha}$ high, $\rho_\alpha$, $w_{\alpha\beta}$ for Te \cite{Woodhouse2013a}; $\eta_\alpha$ high, $\eta_\alpha$ low \cite{Woodhouse2013b}; $t_{\alpha}$ medium, $t_{\alpha}$ low, $U_{\alpha\beta}$ high for Te, $U_{\alpha\beta}$ low for Te, $y_\alpha$ high, $y_\alpha$ medium \cite{Marwede2014}; $\eta_\alpha$ medium \cite{Zuser2011}; $U_{\alpha\beta}$ medium for Te \cite{Candelise2012}; $y_\alpha$ low \cite{Fthenakis2010}. $U_{\alpha\beta}$ values for Cd are assumed to be the same as Te. $w_{\alpha\beta}$ for Cd is 1-$w_{\alpha\beta}$ for Te.

\item \textbf{Si:} $\rho_\alpha$ \cite{Powell2013}; all remaining parameters \cite{Powell2012}. 

\end{tablenotes}  
\end{threeparttable}
\end{table*} 
%%%%%%%%%%%%%%%%%%%%%%%%%%%%%

%%%%%%%%%%%%%%%%%%%%%%%%%%%%%%%
\begin{figure*} %
%\captionsetup{font=scriptsize}

     \begin{center}
        \subfigure[Indium]{%
            \label{fig:in}
            \includegraphics[width=0.33\textwidth]{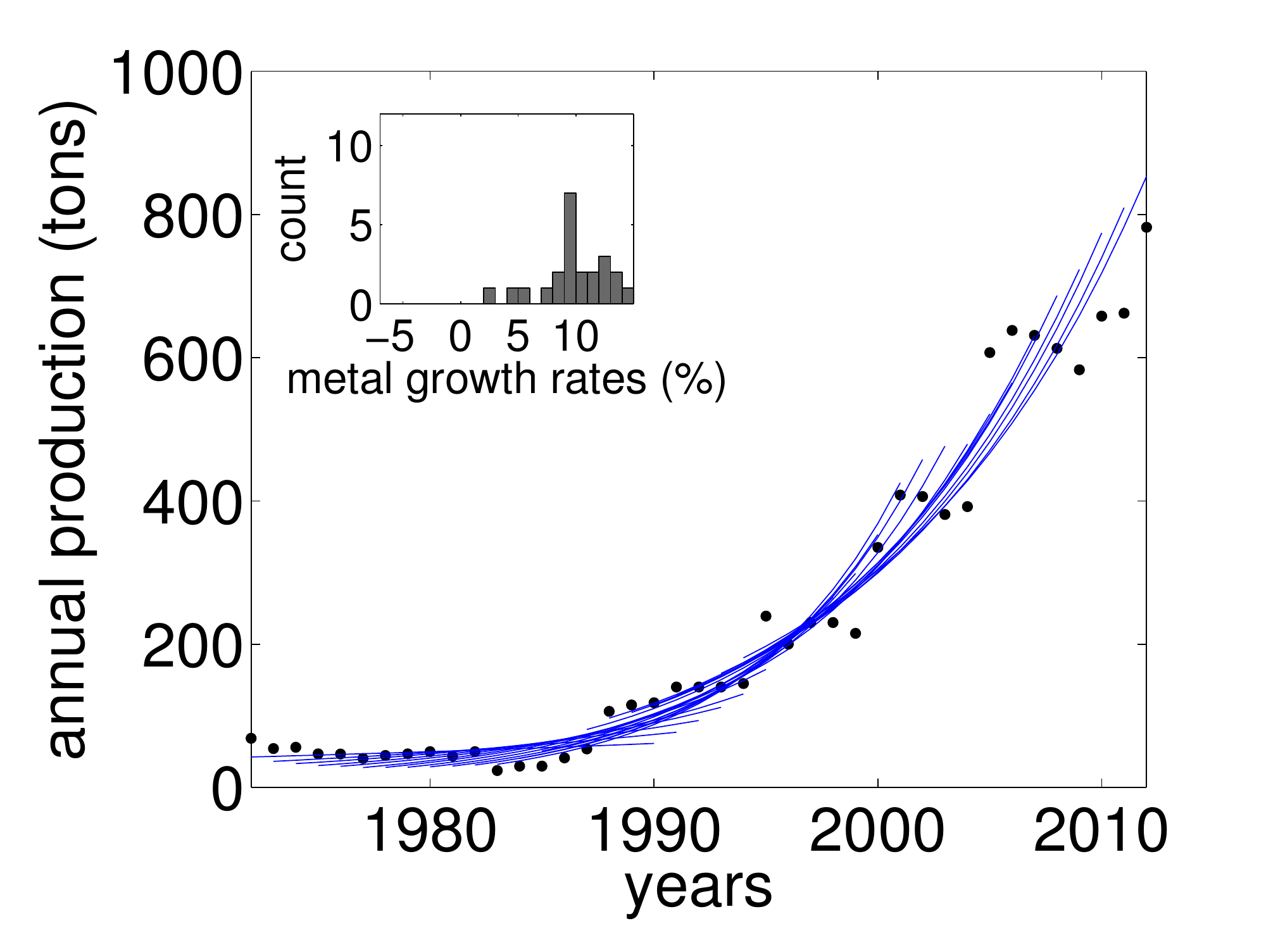}
        }%             
        %  ------- End of the first row ----------------------%
        \subfigure[Gallium]{%
            \label{fig:ga}
            \includegraphics[width=0.33\textwidth]{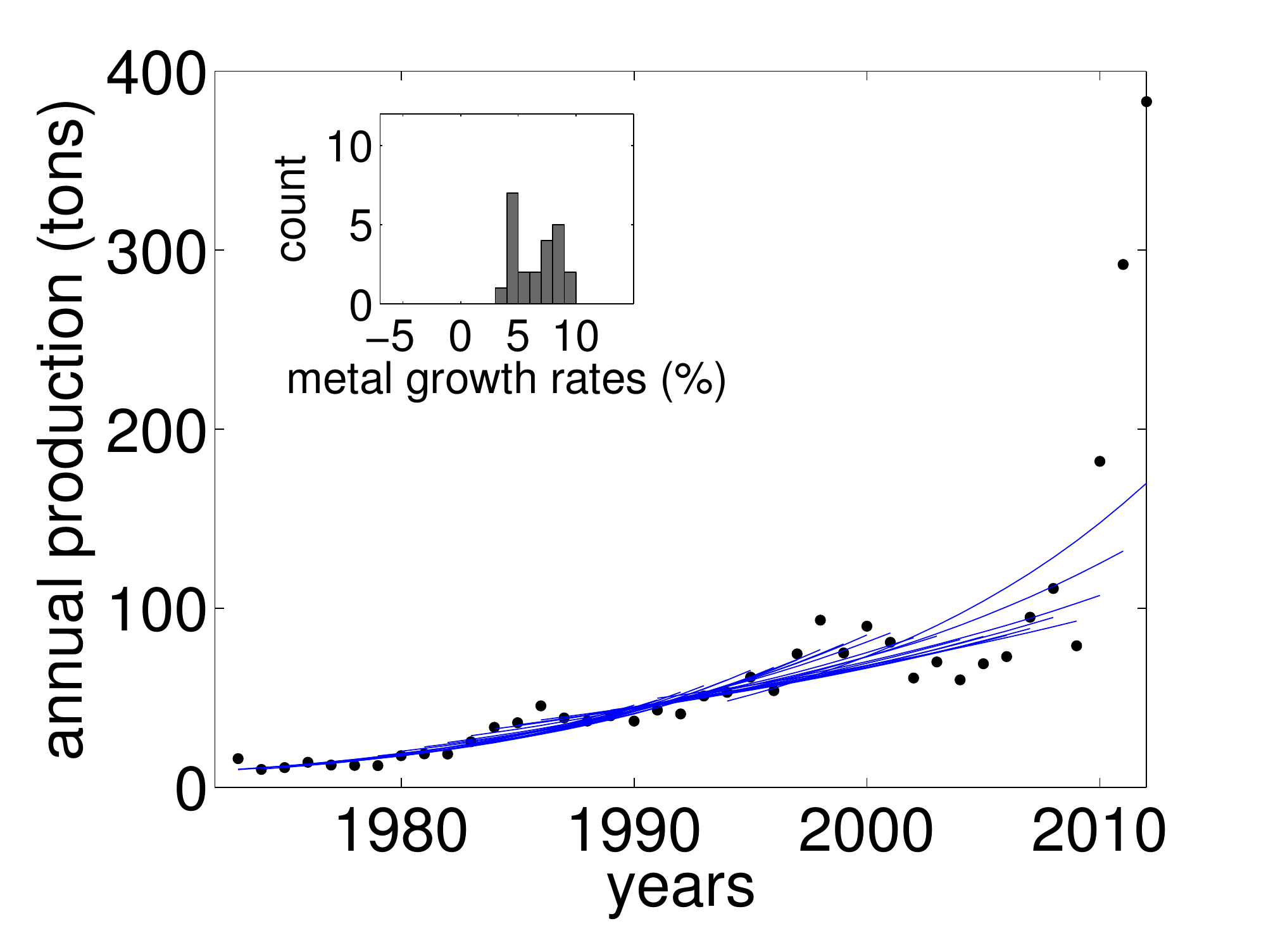}
        }%                           
        %  ------- End of the 2nd row ----------------------%        
        \subfigure[Selenium]{%
           \label{fig:se}
           \includegraphics[width=0.33\textwidth]{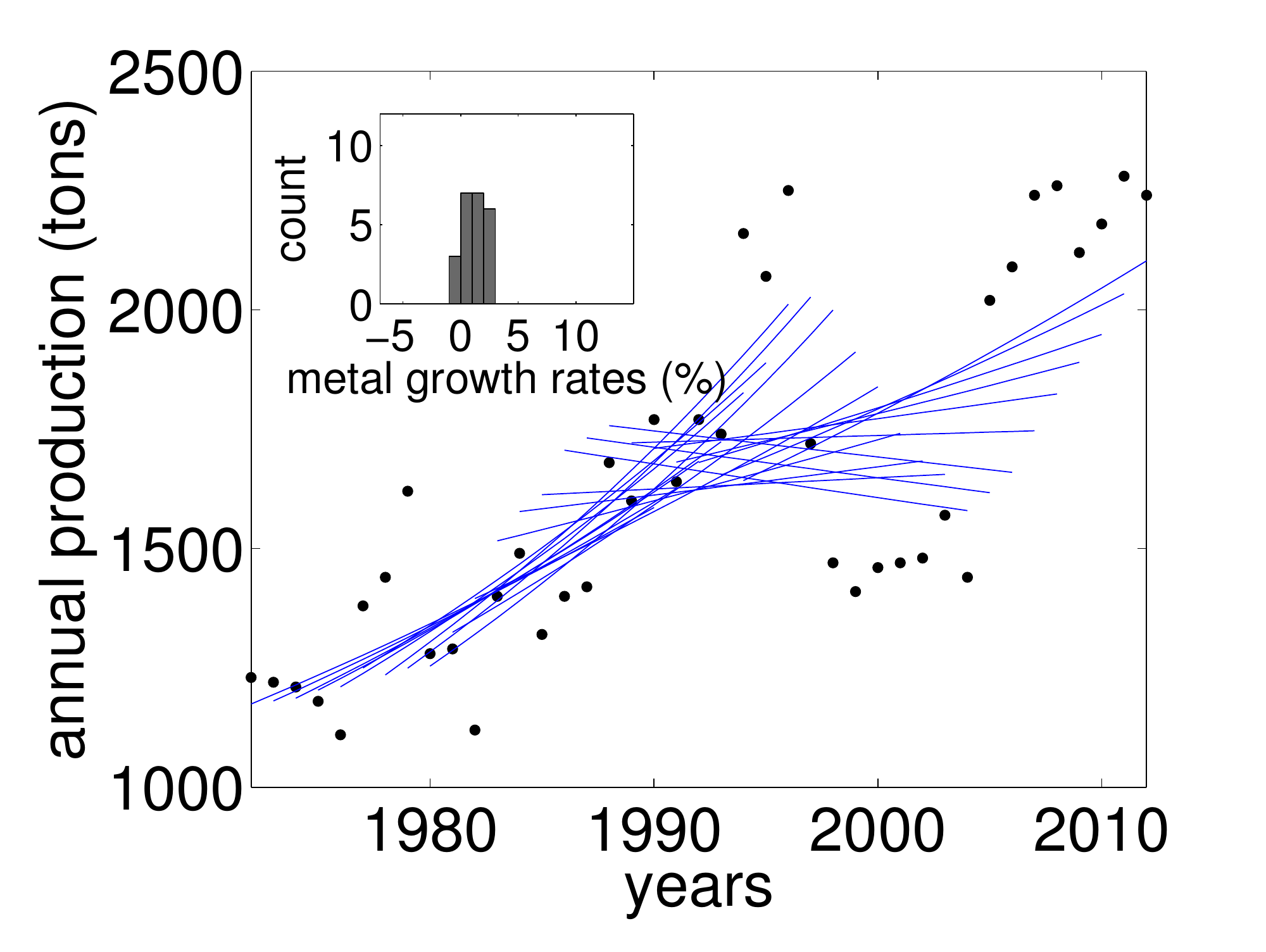}
        }\\        
        \subfigure[Tellurium]{%
            \label{fig:te}
            \includegraphics[width=0.33\textwidth]{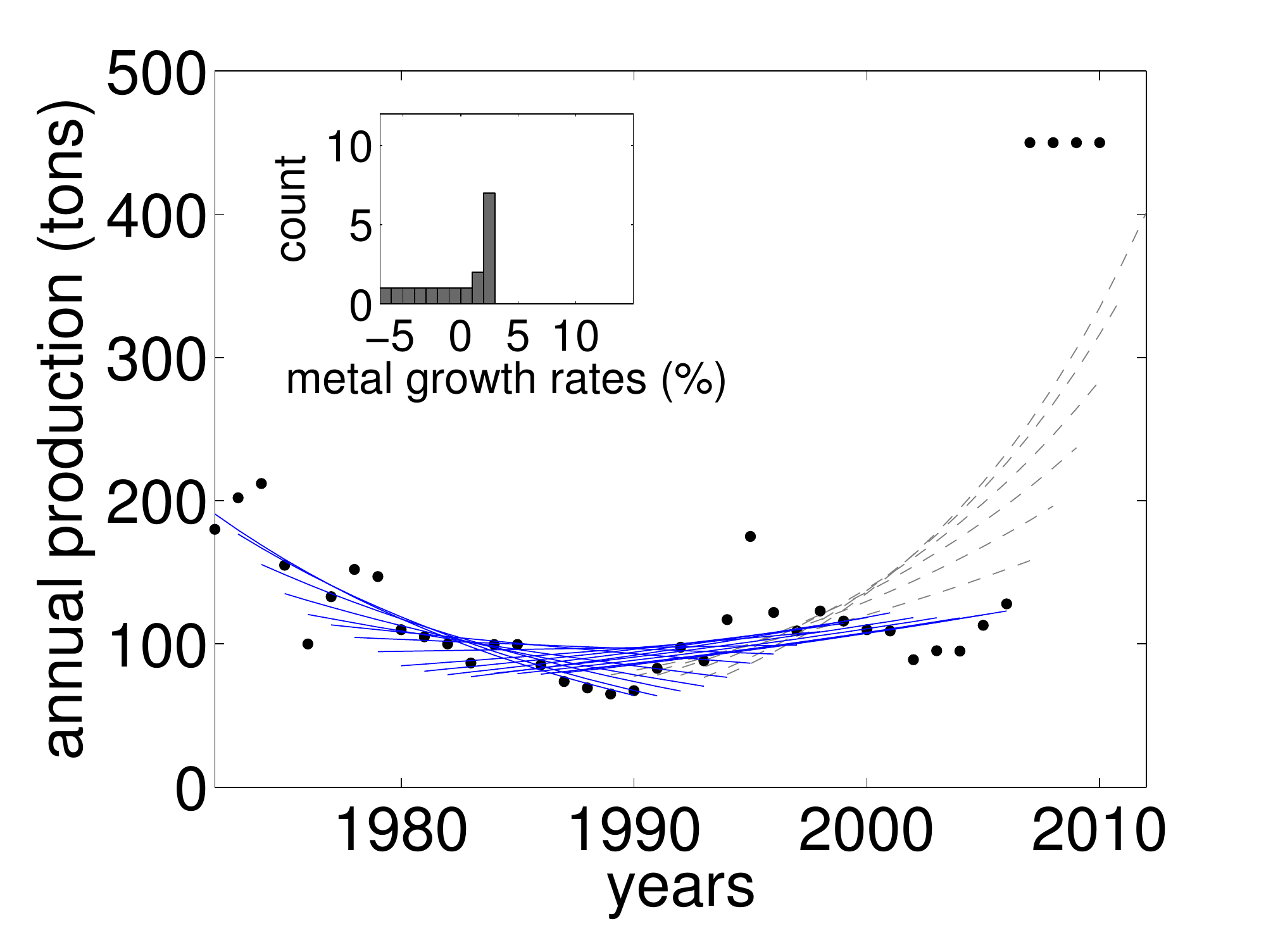}
        }%
        \subfigure[Cadmium]{%
            \label{fig:cd}
            \includegraphics[width=0.33\textwidth]{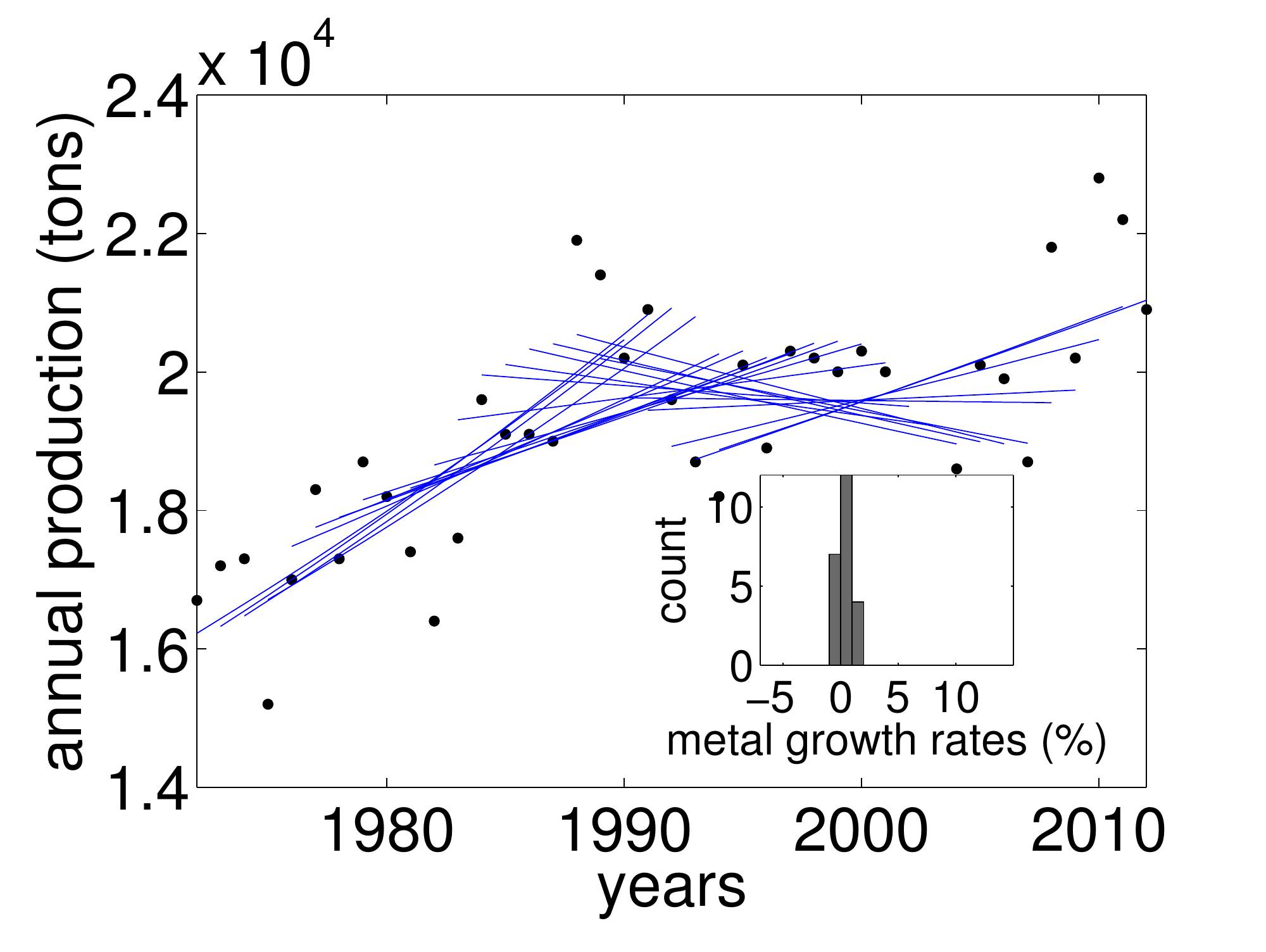}
        }%             
        \subfigure[Silicon]{%
            \label{fig:si}
            \includegraphics[width=0.33\textwidth]{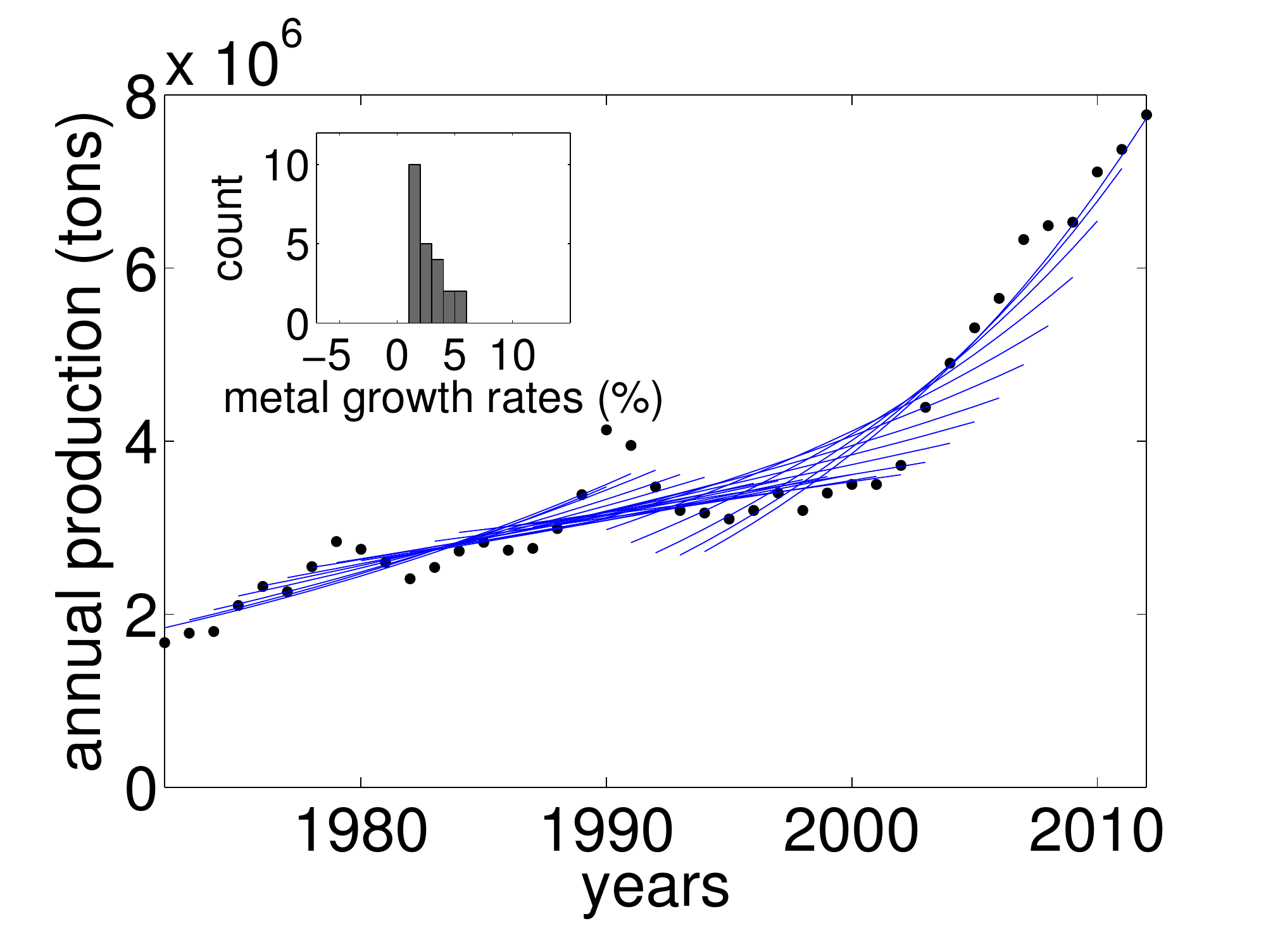}
        }% 
        
        %  ------- End of the 5th row ----------------------%       
      
    \end{center}
\caption{%
Annual production of metals over time, 1972-2012. Black points show the actual production data, while blue lines are obtained by fitting a line to the natural logarithm of the production data (using the least squares method) for each 18-year period in 1972-2012. The slope of the each fitted line represents the annual growth rate for that 18-year period. The inset in each figure is the histogram of the annual growth rates obtained by this curve fitting method. The goodness-of-fit varies substantially across the metals and time periods investigated. The method of reporting tellurium production data changed in 2007, resulting in an arbitrary jump. \cite{private2014} Therefore the Te data for the last 6 years are not taken into account when estimating the growth rates. This is indicated by the gray color used for the last 6 fitted lines.}% Goksin, can you cite personal communication when you make the statement about Te above? GK: Ok,I did.

\label{fig:fig1}
\end{figure*}
%%%%%%%%%%%%%%%%%%%%%%%%%%%%%%%

For each PV metal, we consider a range of estimates for material intensity in 2030. Table \ref{Table2} provides the parameters used to obtain these estimates and the resulting high, medium and low material intensity values. The ranges for material intensity considered are 10-30 t/GW for In, 2-10 t/GW for Ga, and 20-160 t/GW for Se in CIGS; 20-160 t/GW for Te and 20-140 t/GW for Cd in CdTe; and 640-6630 t/GW for Si in c-Si when material losses during manufacturing are considered. The high material intensity estimate corresponds to today's level.

The demand by non-PV end-uses of each metal in 2030 is estimated by using the median of the historical growth rates of that metal over all 18-year periods between 1972 and 2012. To account for the variability in the historical growth rates and the uncertainty regarding the future of the non-PV end-uses of the metal, we also calculate a confidence interval around the median growth of the non-PV end-uses defined by the 1st and 3rd quartiles of the distribution of historical growth rates over all 18-year periods between 1972-2012.

We calculate the growth rate, $r_{\beta}$, required for the metals production in 2012 to reach the required level in 2030 by assuming a constant percentage annual growth rate and using equation (\ref{eq:rbeta}):

\begin{equation}
P_{\beta} = P0_{\beta} \times (1 + r_{\beta}) ^{18}
\label{eq:rbeta}
\end{equation}
where 

\begin{description} [leftmargin=3em,style=nextline, noitemsep]
\item[$P0_{\beta}$] production of metal $\beta$ in 2012 [t/y] (from \cite{USGS2013, USGS2005, USGS2006, USGS2007, USGS2010, USGS2011})
\item[$P_{\beta}$] production of metal $\beta$ in 2030 [t/y] (found in eq. \ref{eq:prod2030})
\end{description}
\vspace*{9pt}

After obtaining the required growth rates, $r_{\beta}$, we compare them to historical growth rates of metals production in order to determine whether the required growth rates have historical precedent. When studying the historical growth rates, we use a large set of metals to obtain a more complete picture of the metals production sector. We obtain the annual global production values for 32 metals for the last 40 years from the U.S. Geological Survey.  \cite{USGS2013, USGS2005, USGS2006, USGS2007, USGS2010, USGS2011} These represent all metals for which continuous yearly production data is available. %Goksin please check. GK: GK: Correct. We don?t use B, Ti and Zr because the reported production includes concentrates (oxides etc) of these metals and reported in gross weight. I also didn?t use He although the production is reported.

We study material resources at the purity grade reported by the US Geological Survey. (See Table S1 in the Supplementary Information.\dag) We note that byproduct metals such as Te are generally tracked at higher levels of purity than primary metals such as Si, since only the refined byproduct is globally traded. Because of this, we carry out an additional analysis on metallurgical grade Si, a higher purity form that is the precursor to most (97\%) Si used in solar cells \cite{Yue2014,Yue2014corr,Safarian2012}, to see whether this partially-refined material with smaller production scale is able to support deployment of Si-based PV (Supplementary Information\dag). This analysis also limits the raw Si resource, since currently metallurgical grade Si is produced more selectively from silica deposits with relatively low starting level of impurities. \cite{Goetze2014} We note that data on MG-Si is limited to the period from 1990 - 2012, and that data prior to 2004 excludes production by China, further limiting the number of observations. To maintain consistency with other metals, in figures \ref{fig:fig1}-\ref{fig:fig4} we use total production of Si given in the USGS data, for which a full 40 year history of most recent production data is available.

We calculate the historical annual growth rates for each metal for all overlapping 18-year periods between 1972-2012. Annual growth rates are calculated based on 18-year time horizons to match the time horizon of the metals growth projections considered (2012-2030). Because we are interested in growth rates that are sustained over all possible 18-year periods, we measure the growth rates over overlapping periods rather than disjoint periods. The average annual growth rate of metals production over each 18-year period is estimated by fitting a straight line to the natural logarithm of the production over time using the least-squares method (figure \ref{fig:fig1}). This is not meant to be a high-fidelity model and we emphasize that the goodness-of-fit varies substantially across the metals and 18-year time periods studied. This level of fidelity is appropriate for answering the following question: If we approximate past and future growth in metals production as following an exponential trend over an 18-year period, how do future required growth rates compare to those observed in the past?  %This results in the growth rates in middle years being included in more 18 year periods sampled than years at either the beginning or end of the period 1972-2012. %However this has minimal effect on the distribution as there is no evidence that these middle years show a systematically higher or lower growth rate than other year. %This leads to growth rates that are not independent from each other; however, these growth rates are only used as benchmarks for the required growth rates for the future.

%%%%%%%%%%%%%%%%%%%%%%%%%%%%%%%%%%%%%%%%%%%%%%%%%%%%%%%%%%%%%%
\section{Results and Discussion} \label{sec:Results}
%\input{results}

% POINT: Introduce the results section
In this section, we first present historical growth rates in the production of metals. Next, we show the growth rates required for the PV metals to reach various projected annual PV installation levels in 2030, and compare these to historical growth rates. Third, we briefly discuss constraints on scaling up the production of byproduct metals based on the production levels of their host metals, as well as the estimated metals reserves. 

\subsection{Historical Growth Rates}

% POINT: Discuss the historical growth for PV metals, by first showing the fitted lines
Figure \ref{fig:fig1} shows the annual production values for a set of PV metals over time %, and the annual growth rates observed by these metals over all 18-year periods in the last forty years 
(1972-2012). The inset in each plot shows a histogram of the annual growth rates for the corresponding metal over all 18 year periods. As can be seen in figure \ref{fig:fig1}, the variability in annual growth rates differs across metals but the distribution of growth rates is constrained to a fairly narrow range, falling below 10\% growth per year for these metals with the exception of In. 

\begin{figure} [!h]%[e]
     \begin{center}
%\subfigure[Historical growth rates over time for PV metals of interest]{%
            \includegraphics[width=0.45\textwidth]{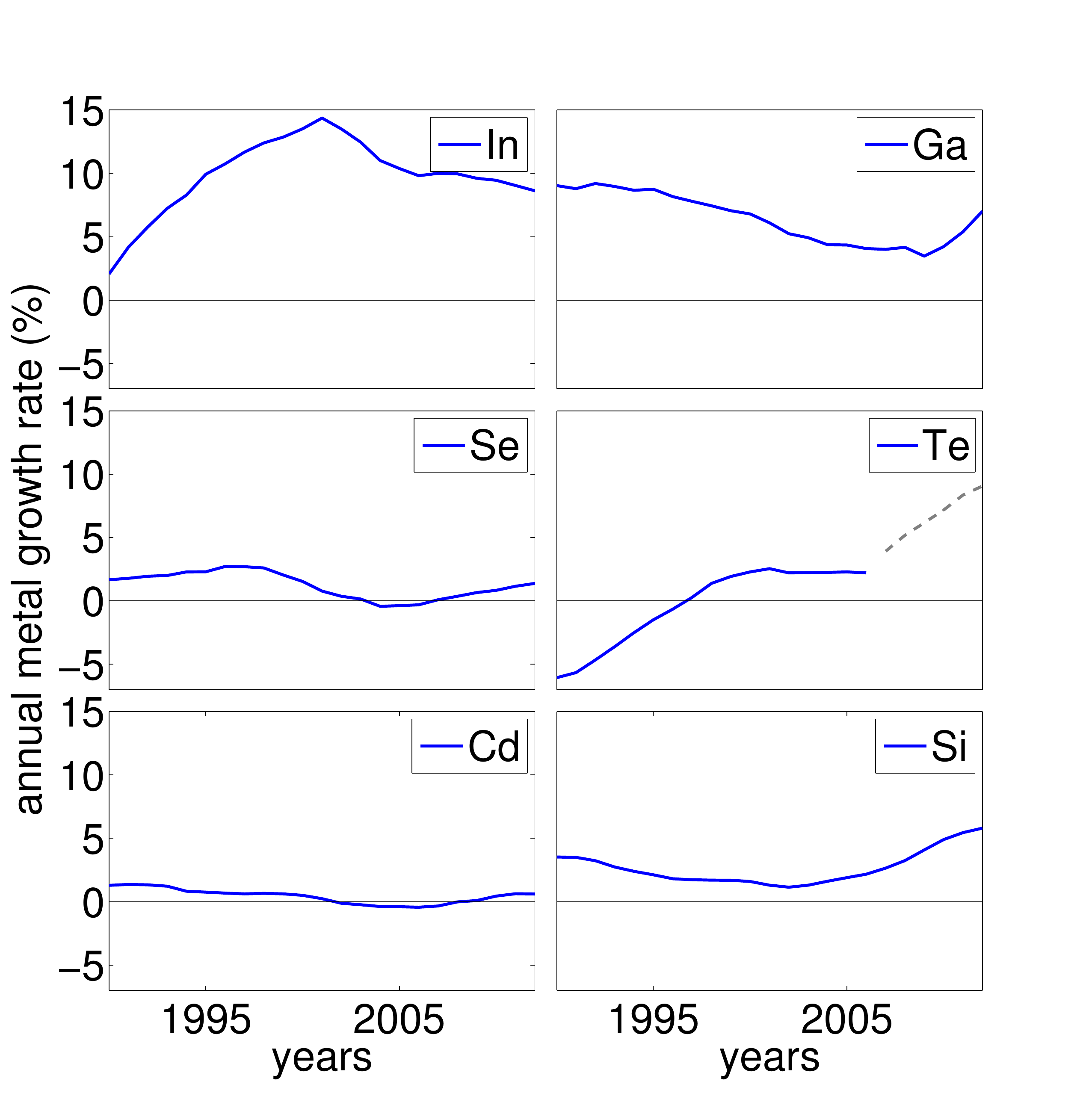}
%        }%

\end{center}
\caption{%
Historical growth rates over time are shown for the metals of interest. Growth rates are calculated and plotted in a backward looking manner: The growth rate corresponding to a year is calculated using the production values from the previous 18 years. Reporting of the Te production data changed in 2007, \cite{private2014} therefore the data for the last 6 years are not taken into account when estimating the growth rates. 
}%

\label{fig:fig2}
\end{figure}

% POINT: Show the evolution of growth rates for PV metals over time. Highlight the differences between metals.
The change in growth rates over time are different for each PV metal (figure \ref{fig:fig2}). In and Ga have experienced growth rates that are mostly above 5\% per year, which is high compared to the other four metals. In, Ga, and Te growth rates have changed significantly over time, unlike Se and Cd rates, which have fluctuated within a small range between -1\% and 3\% per year. Si growth rate has also been lower and more stable compared to In, Ga, and Te, and recently increased to 5\% per year.  

%%%%%%%%%%%%%%%%%%%%%%%%%%%%%%%
%%%%%%%%%%%%%%%%%%%%%%%%%%%%%%%

\begin{figure} [!h]

     \begin{center}
        \subfigure[Histogram of historical growth rates]{%
            \label{fig:histog}
            \includegraphics[height=0.33\textwidth]{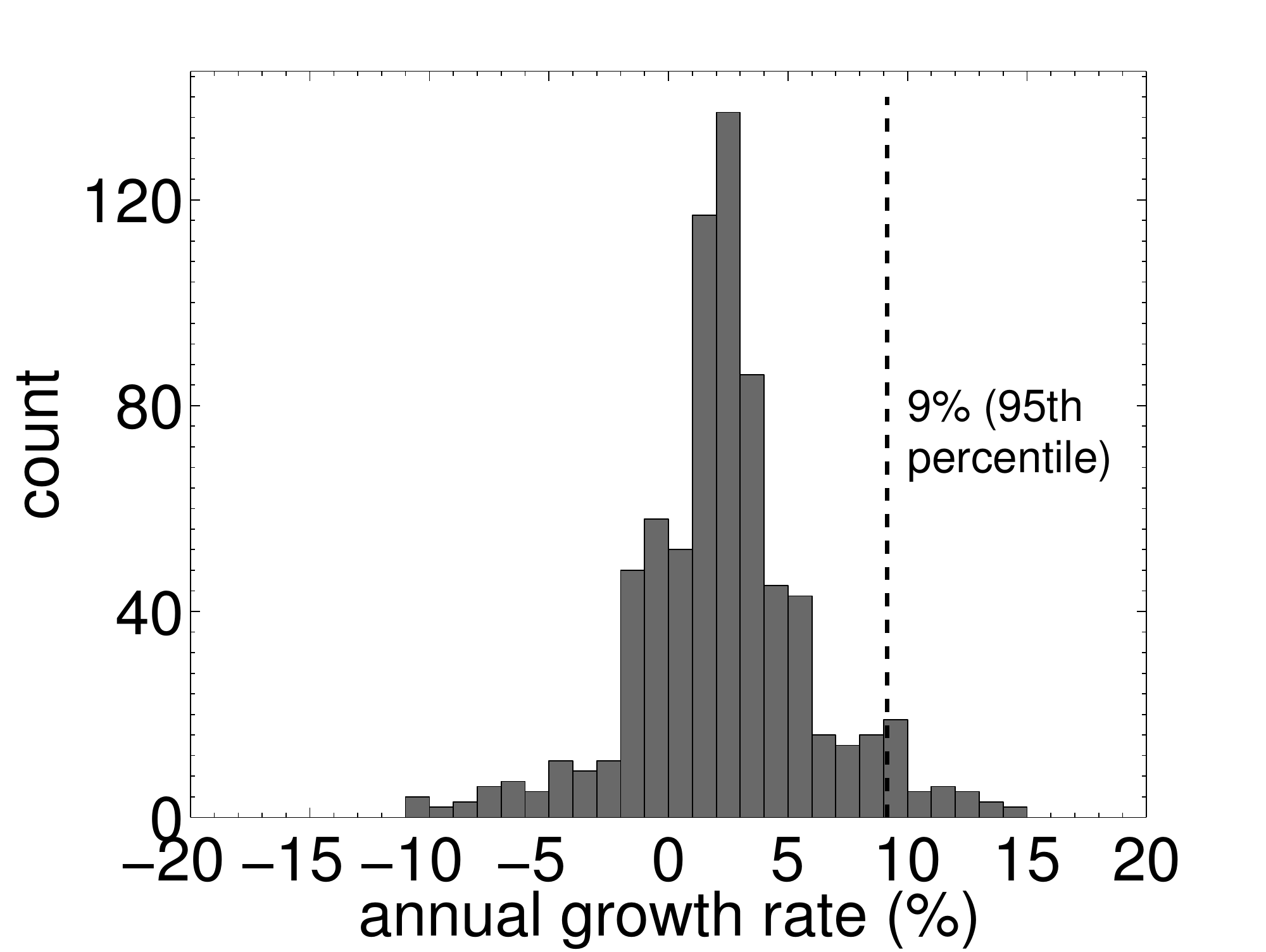}
        }%
        \\
        \subfigure[Historical growth rates over time]{%
            \label{fig:histTime}
            \includegraphics[height=0.33\textwidth]{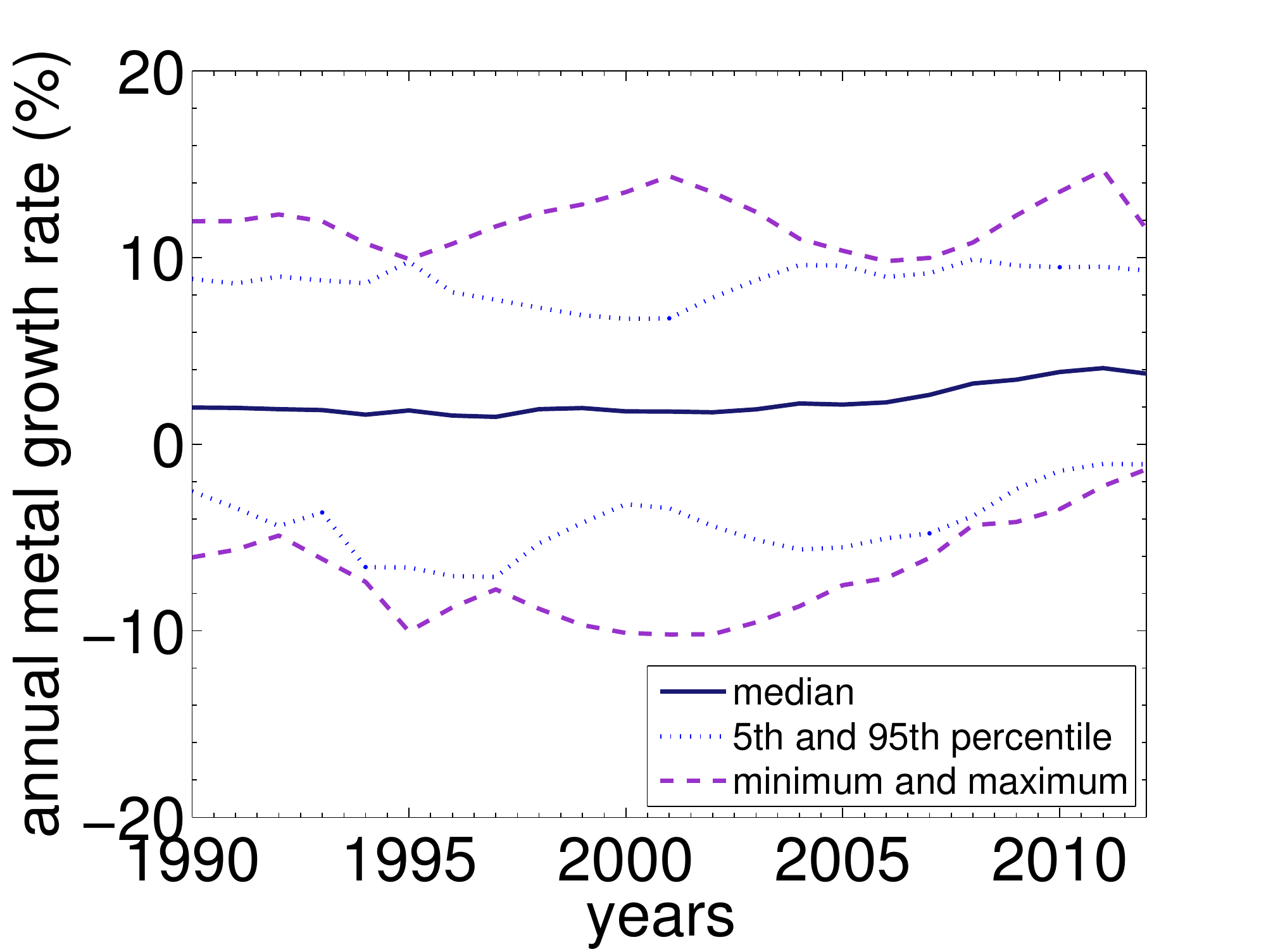}
        }%      

    \end{center}
\caption{%
\textbf{(a)}: Histogram shows the distribution of the historical annual growth rates of production of 32 metals observed in 1972-2012 over 18-year periods. Growth rates are calculated by fitting lines to the natural logarithm of the production values in each of the 18-year periods in 1972-2012. The median 18-year average annual growth rate is 2.3\%.  \textbf{(b)}: 18-year average annual growth rates in metals production have been almost constant over time for the 32 metals studied. Annual growth rates are backward looking: they are calculated using the production values from the previous 18 years. The solid midline is the median of the growth rates of 32 metals for each year. The blue dotted lines show the 5th and 95th percentiles. The dashed purple lines show the minimum and the maximum growth rates observed. Note: The data coming from the last 6 years of Te production are excluded from both panel (a) and panel (b) due to a change in the reporting of the Te production data in 2007. \cite{private2014} %\textbf{(c)}: Historical growth rates over time are shown for the metals of interest. 
}%

\label{fig:fig3}
\end{figure}

% POINT: Historical rates of many metals are calculated and aggregated to obtain a distribution. 
To gain a broader picture of the metals production industry, we also obtain the growth rates for the 32 metals available in the USGS database (figure \ref{fig:fig3}). %Goksin - does the journal allow an appendix or SI? We may need to list the other metals somewhere. GK: I listed them in SI.
Figure \ref{fig:histog} shows the histogram of the aggregated growth rates observed by the set of 32 metals over all 18-year periods in 1972-2012. We see that the median growth rate is 2.3\% per year. A growth rate of 9\% per year at the 95th percentile of the aggregated growth rate distribution is marked with a vertical dashed line in figure \ref{fig:histog}. We interpret 9\% per year as an upper end of business-as-usual growth. If a growth rate of 9\% per year is sustained over an 18-year period, the annual production will increase by almost a factor of 5 over this period. Also important to our analysis is the maximum growth rate that has been sustained over an 18-year period, which is 14.7\% per year.  

%%%%%%%%%%%%%%%%%%%%%%%%%%%%%%%
%%%%%%%%%%%%%%%%%%%%%%%%%%%%%%%

% POINT: The aggregated growth rates over time are observed for trends.
Figure \ref{fig:histTime} shows how the historical growth rates change over time. We observe that the median 18-year average growth rate has been mostly stable over time. An upward trend is observed in the median as well as the interval between the 5th and 95th percentiles after 2005. Even with this upward trend in recent years, the growth rates have been stable, and the median growth rate stayed below 5\% per year.

Using overlapping time periods to obtain the 18-year average annual growth rates places greater weight on years falling in the middle of the time span considered (1972-2012). However we note that this does not introduce a bias in the histogram shown in figure \ref{fig:histog}, as the annual production growth rates across the aggregated set of metals trend neither up nor down over the period considered (1972-2012), as shown in figure S1 in Supplementary Information.\dag~ %Goksin - you could add a few supporting plots to the SI to support this. Also, you could add to the SI plots showing all production data for all metals.

\subsection{Comparison of Projected and Historical Growth Rates}

\begin{figure*} %[!h]
     \begin{center}
        \subfigure[Indium, $n_{In}$ = 10\% ]{%
            \label{fig:in2}
            \includegraphics[width=0.32\textwidth]{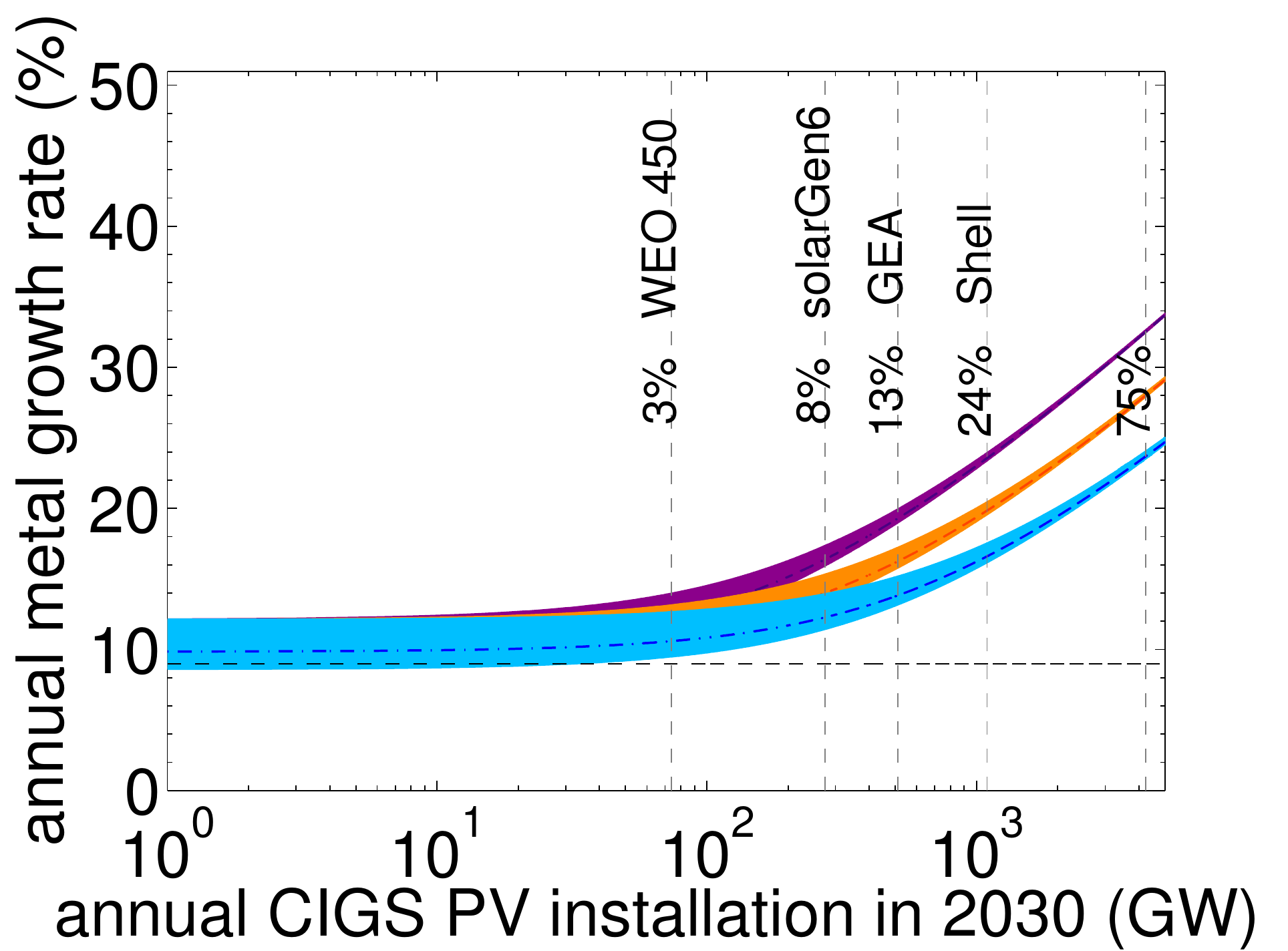}
        }%        
         \subfigure[Gallium, $n_{Ga}$ = 6.8\% ]{%
           \label{fig:ga2}
           \includegraphics[width=0.32\textwidth]{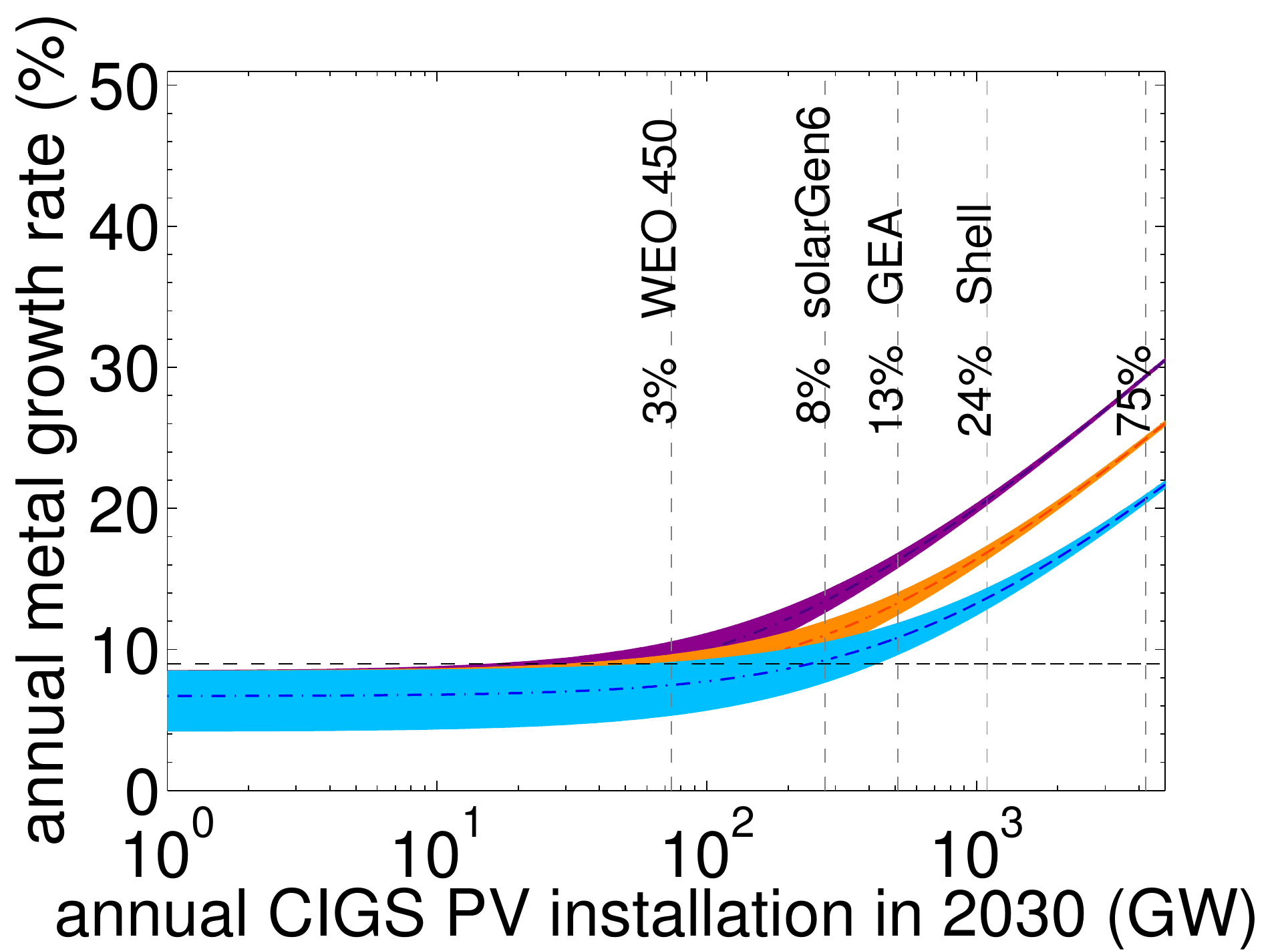}
        }%
         \subfigure[Selenium, $n_{Se}$ = 1.4\% ]{%
           \label{fig:se2}
           \includegraphics[width=0.32\textwidth]{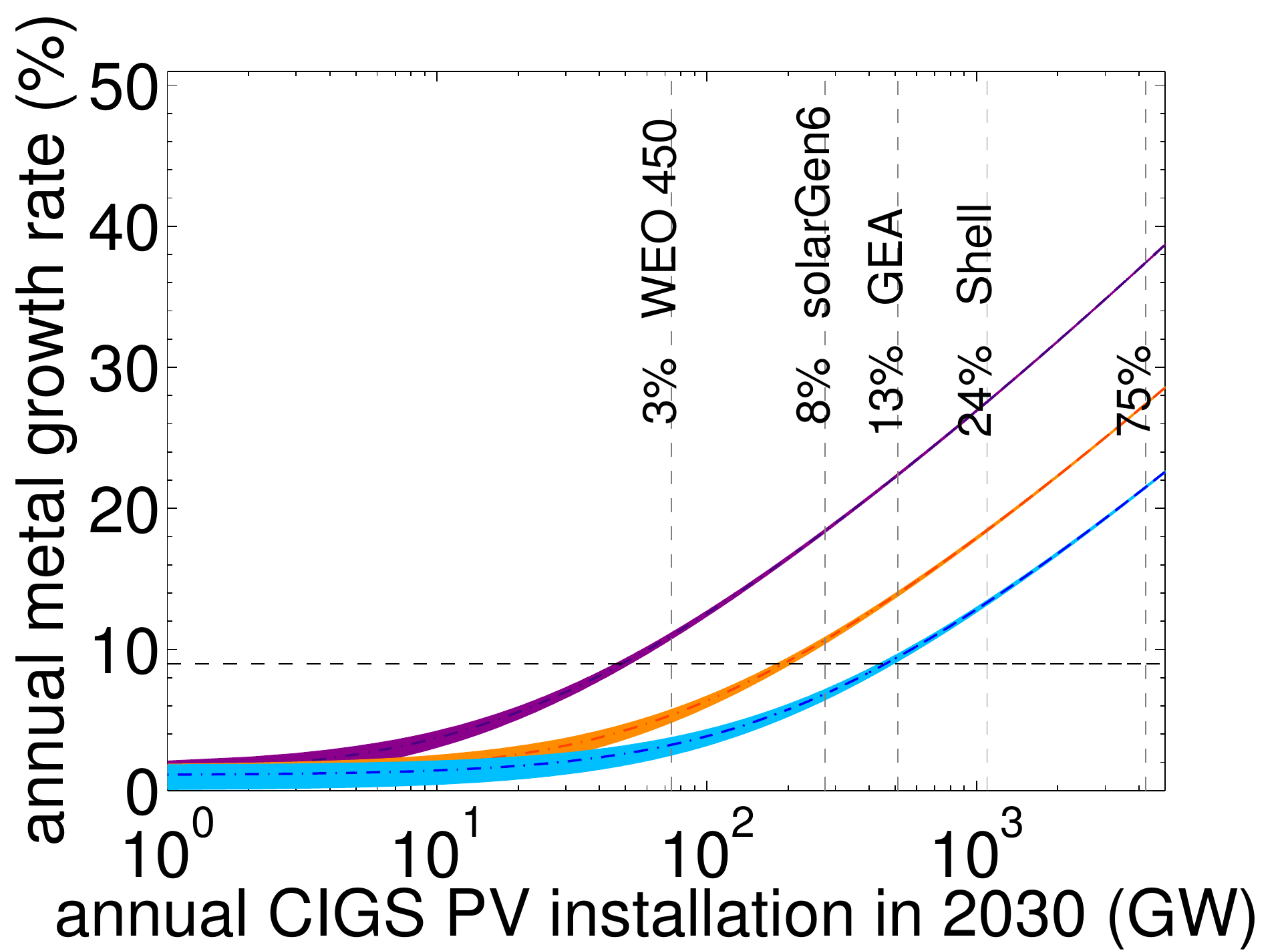}
        }%
        \\
 			 %  ------- End of the first row ----------------------%
        \subfigure[Tellurium, $n_{Te}$ = 1.4\% ]{%
            \label{fig:te2}
            \includegraphics[width=0.32\textwidth]{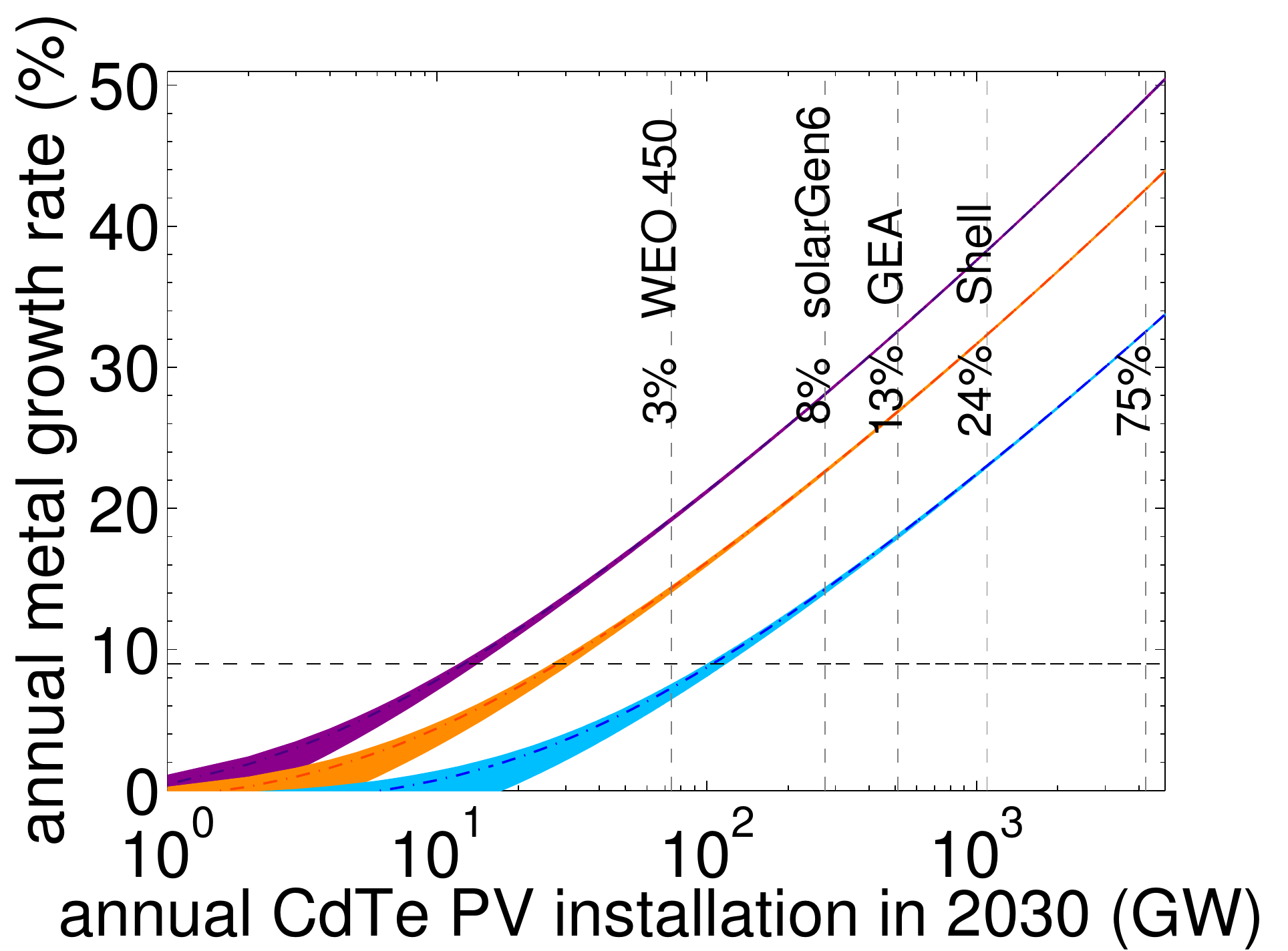}
        }%
        \subfigure[Cadmium, $n_{Cd}$ = 0.6\% ]{%
            \label{fig:cd2}
            \includegraphics[width=0.32\textwidth]{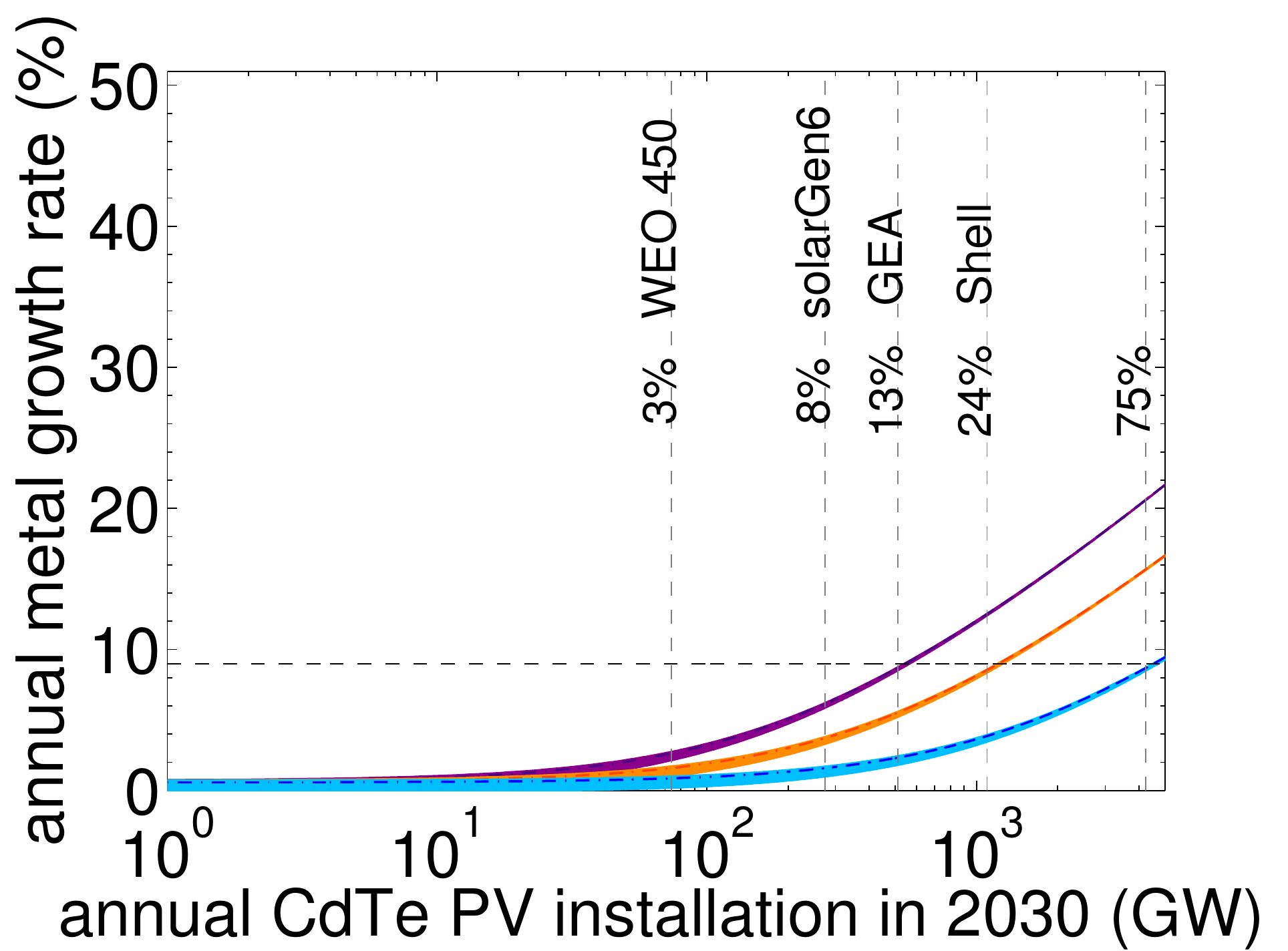}
        }%
        \subfigure[Silicon, $n_{Si}$ = 2.2\% ]{%
            \label{fig:si2}
            \includegraphics[width=0.32\textwidth]{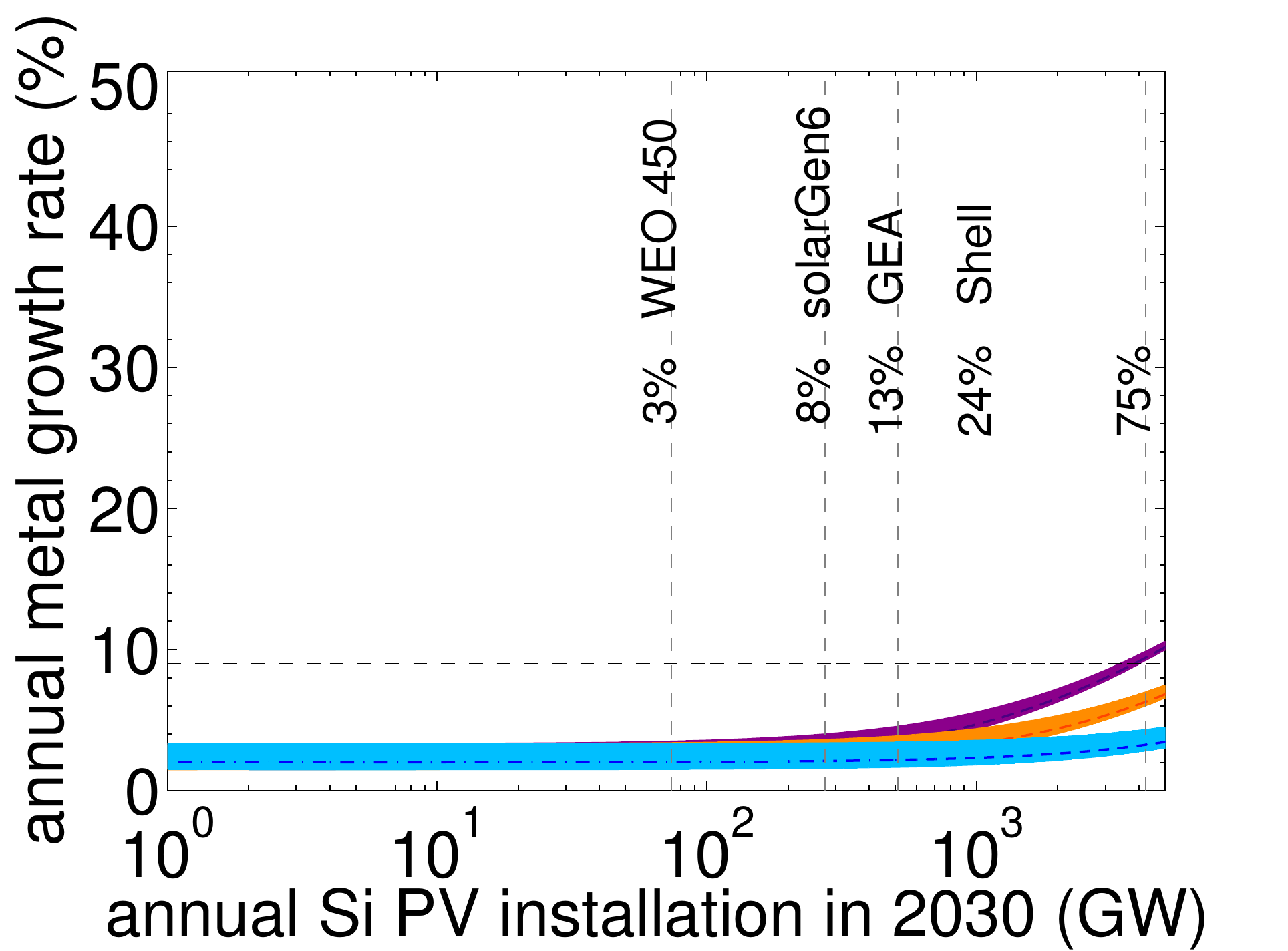}
        }%      
        \\  
        \subfigure{%
            \label{fig:legend}
            \includegraphics[width=0.55\textwidth]{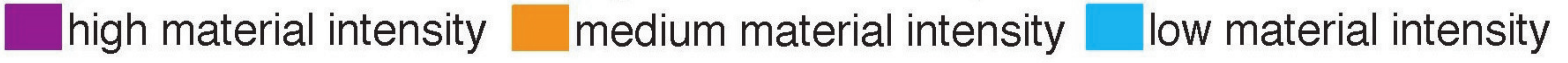}
        }

    \end{center}
    \caption{%
Required growth rates for metals production to reach a range of annual PV installation levels in 2030. The bands with different colors show the required growth rates for different levels of material intensities given in table \ref{Table2}. The lower and upper ends of each band are obtained by assuming that the non-PV end-uses grow at rates equal to the 1st and 3rd quartiles, respectively, of the historical growth rate distribution of that metal over each 18-year period between 1972-2012. % how about doing the wedges for the 5-95th percentiles instead of quartiles for consistency.
The median of the 18-year average growth rates observed between 1972-2012 for each metal, ($n$), is shown below each plot. The vertical lines indicate the assumed annual installation level for the PV technology corresponding to each energy scenario. The percentage on the left of the scenario names indicate the fraction of global electricity generation coming from PV. The energy scenarios originally report only the cumulative PV installations. By assuming a constant percent annual growth rate, we calculated the annual installation level in 2030. The horizontal line at 9\% growth rate corresponds to the 95th percentile of the historical growth rates for all 32 metals as shown in figure \ref{fig:fig3}.
     }%

\label{fig:fig4}
\end{figure*}

% POINT: Introduce the "wedges" figure, and introduce which material intensity case & PV scenario will be discussed.  
Figure \ref{fig:fig4} shows the annual growth rates required for the production of PV metals to meet the demand of a wide range of annual PV installation levels in 2030. The lower and upper ends of each colored band in figure \ref{fig:fig4} are based on growth in non-PV end-uses at rates defined by the 1st and 3rd quartiles of the distribution of their historical 18-year average growth rates. 

% POINT: Talk about the solar Gen 6 scenario & other assumptions used to obtain the annual installation level.
When explaining the results, we focus on the medium material intensity case and the annual PV installation level corresponding to the ``Solar Generation 6, paradigm shift''\cite{EPIA2011} scenario. %The medium material intensity is approximately equal to half of the high material intensity for all of the five PV metals except for selenium, in which case it is about one fourth of the high material intensity. 
The Solar Generation 6 \cite{EPIA2011} scenario projects a relatively modest growth in PV installations. In this scenario, the cumulative installed capacity reaches 1850 GW in 2030, \cite{EPIA2011} and generates around 2430 TWh in a year assuming an average capacity factor of 15\%. If the total annual global electricity generation in 2030 is 30000 TWh, \cite{IEA2012} then PV supplies around 8\% of the world's electricity in this scenario. When we assume that PV installation grows at a constant percentage annual growth rate starting from about 100 GW cumulative installed capacity in 2012, the annual PV installations in 2030 would be 275 GW to reach the 1850 GW cumulative installed capacity in 2030. In figure \ref{fig:fig4}, the vertical line marked with the ``8\% SolarGen6'' corresponds to this annual installation level, and 8\% refers to the portion of the global electricity provided by PV. 

% POINT: give results for CIGS metals growth rates
If CIGS provides all of the 275 GW annual installations projected by the Solar Generation 6 \cite{EPIA2011} scenario in 2030, the required growth rate for In is approximately 14\% per year for the medium intensity case, as shown in figure \ref{fig:in2}. This rate is almost unprecedented considering that the highest growth rate that has been observed historically by a large group of metals is 14.7\% (as shown in figure \ref{fig:histog}). 14\% per year is also high compared to In's recent growth rates, where In production has been growing at a rate lower than 10\% per year (figure \ref{fig:fig2}). A growth rate of 14\% per year means that In production increases from 780 t/y in 2012 to 8250 t/y in 2030, which is over a factor of 10 increase. In this scenario, the annual Ga and Se production each needs to grow at 11\% per year, as shown in figure \ref{fig:ga2} and figure \ref{fig:se2}. These growth rates are greater than the majority of the historical growth rates experienced by all metals (figure \ref{fig:histog}). For Se, the projected 11\% per year is significantly higher than the growth rates that Se has experienced in the last forty years (figure \ref{fig:fig2}), and corresponds to an increase in annual production of 2240 t/y in 2012 to 14660 t/y in 2030. On the other hand, the required growth rate for Ga, 11\% per year, is slightly above what Ga has been experiencing in the recent years (figure \ref{fig:fig2}). This rate means that Ga production increases from 380 t/y in 2012 to 2490 t/y in 2030.

%Another important piece of information provided by Fig. \ref{fig:fig4} is the maximum level of PV deployment that would prevent the growth rate of metals production to exceed the historically observed rates by all of the 32 metals. 
If we constrain metals growth rates to the maximum historical rate of 14.7\%, the annual CIGS deployment levels in 2030 would be limited by In to 340 GW. %In this case, the cumulative PV installation would be about 2180 GW and PV would provide 9.5\% of the global electricity generation in 2030. 
Se and Ga would allow the annual CIGS deployment to be up to 580 GW and 700 GW, respectively, if CIGS were not limited by In.

% POINT: give results for CdTe metals growth rates
If CdTe provides 8\% of the global electricity generation in 2030 corresponding to the Solar Generation 6 \cite{EPIA2011} scenario, in which case the annual CdTe installation in 2030 is 275 GW, Te production needs to grow at 23\% per year for the medium material intensity case as shown in figure \ref{fig:te2}. 23\% per year is significantly higher than the highest historical growth rate observed for all of the 32 metals (figure \ref{fig:histog}). 23\% per year growth rate corresponds to a more than fortyfold increase in the annual Te production - from 500 t/y in 2012 to 20760 t/y in 2030. Cd, on the other hand, requires only 4\% per year growth rate in this scenario (figure \ref{fig:cd2}) - an increase from 20900 t/y to 42340 t/y. Historically, Cd production has been growing at very low rates and even decreased over sustained time periods as can be observed in the negative rates in figure \ref{fig:fig2}. 4\% growth per year is relatively low compared to the required growth rates of other byproduct metals.

The maximum CdTe deployment in 2030 would be determined by Te, if Te growth rate does not exceed the maximum historically observed growth rate (14.7\% per year) observed by all of the 32 metals. In this case, the annual CdTe deployment in 2030 would be limited to 80 GW. %Accordingly, the cumulative PV installation would be 760 GW, leading to 3.3\% of global electricity to be obtained from PV. 
Cd would allow up 3600 GW annual CdTe deployment, if there were no constraints imposed by Te.

% POINT: explain why the In, Ga AND Te plots look different: historical rates and % of metal dedicated to PV use are different.
The median historical growth rates for In and Ga are $n_{In}$ = 10\% and $n_{Ga}$ = 6.8\%, are on the higher end of the historical growth rates of all metals. Since we project the growth of non-PV end-uses based on the historical growth rates and the share of non-PV end-uses is very high compared to PV uses, the non-PV demand for both In and Ga is projected to be high. In comparison, a larger fraction (40\%) of Te is used for PV compared to In and Ga, which have only up to 5\% of their production dedicated to PV uses. For this reason, the required Te growth rates are more directly related to the level of PV installations than CIGS metals are as seen in figure \ref{fig:fig4}.

% POINT: give results for silicon growth rates
Instead of CIGS or CdTe, if all of the 275 GW annual PV installation comes from c-Si, Si production needs to increase only by 2.5\% per year (figure \ref{fig:si2}). This growth rate is close to the median historical growth rate observed for all metals, 2.3\% per year. Unlike byproduct metals, the increasing PV deployment does not cause much increase in the required growth rates for Si. This is mainly due to the fact that PV constitutes only a tiny fraction of Si's end-uses. For the medium material intensity case, the required growth rate for Si does not exceed 5\% per year up to 1000 GW of annual deployment. For all annual deployment levels explored in this analysis (up to 6000 GW per year in 2030, supplying 100\% of forecasted electricity consumption) the required growth rates for Si stay within the range of historical growth rates observed for all metals. (We estimate that silver production for use in contacts for c-Si cells can supply high levels of c-Si PV deployment (up to 80\% of global electricity by c-Si in 2030) without exceeding historical growth rates. See the Supplementary Information for details.\dag~Silver might also be replaced with other materials.) \cite{Goodrich2013, Kim2013, urRehman2014} We find the same results using only metallurgical grade Si (see Methods section) as the basis for Si growth rate measurements, rather than all Si. Applying the same analysis to this alternative measure of useable Si production, in the Supplementary Information\dag, we find that required growth rates remain within the historical range with 100\% of global electricity supplied by Si-based solar cells.

% POINT: give results for silicon growth rates: scale of production is much larger for Si than other metals.
It is worth noting that although the growth rates are lower for Si, the increase in the amount of annual Si production from 2012 to 2030 is two to three orders of magnitude higher compared to other PV metals (In, Ga, Te, Se) because it is produced at a much larger scale. A growth rate  of 2.5\% per year correspond to an increase in Si production from 7.8 millions t/y in 2012 to 12.2 million t/y in 2030.  

% POINT: introduce the concern about scaling up production of byproducts and what this section will do: How much can the production of byproducts can be scaled up if all of the available byproduct is extracted from their common host mineral?

%%%%%%%%%%%%%%%%%%%%%%%%%%%%%%%%
%%%%%%%%%%%%%%%%%%%%%%%%%%%%%%%%

\subsection{Discussion of Constraints on Metals Production Growth}

In, Ga, Se, Te, and Cd have low crustal abundances and are extracted economically today only as byproducts of other `host' metals. However a significant quantity of byproduct metal is never extracted from the mined ore. Below we briefly discuss the scalability potential of byproduct metals if they were recovered with 100\% efficiency from the mined mineral at today's production levels of host metals, and compare these production levels to those required to meet 8\% of global electricity (corresponding to the Solar Generation 6 scenario.) \cite{EPIA2011} We also compare the projected metals production requirements to estimates of global metals reserves. We note, however, that the reserves estimates are revised over time to reflect newly identified mineable deposits, and therefore should not be considered fixed constraints on metals production. \cite{Rogner1997, Kesler2010}

We find that for a scenario in which 8\% of the global electricity in 2030 is provided by PV (CIGS, CdTe, or c-Si), and under the assumption of medium material intensity, the required levels of annual production for In, Te, and Se exceed the estimated potential production levels for today by large amounts. The required annual Te production in 2030 also exceeds the Te reserves, \cite{USGS2014} while the required annual In production in 2030 approaches the estimated In reserves. The cumulative production by 2030 would far exceed the reserves for In and Te. Ga and Si production are less constrained as discussed further below.

The amount of annually recoverable In is 1350 t/y based on the average In content of the zinc ore, sphalerite (ZnS),\cite{Bradshaw2013} and the annual zinc production of 13.5 million tons in 2012. \cite{USGS2013} The required annual In production in 2030 to meet 8\% of electricity demand ({\raise.17ex\hbox{$\scriptstyle\sim$}}8300 t/y) is about 6 times the annually recoverable In (1350 t/y), and close to the estimated global In reserves (11000 t). \cite{USGS2008} 

% We need to be careful in how we are referring to these energy scenarios. We make an assumption about the capacity factor and the projected electricity demand in order to arrive at our estimated percent of electricity met by PV. We can say that the percent of electricity met by PV corresponds to a given energy scenario, but we can't say that these are the published energy scenarios. I corrected this in several places but please keep an eye out for it.

The potential Te and Se production can be estimated to be around 1430 t/y and 5500 t/y, respectively, based on the average Te and Se content of the anode in the electrolytic copper refineries \cite{Green2006} and the global electrolytic copper refinery production in 2011. \cite{USGS2011Cu} The required production for Te to meet 8\% of electricity demand in 2030 ({\raise.17ex\hbox{$\scriptstyle\sim$}}20800 t/y) is an order of magnitude larger than the potential production ({\raise.17ex\hbox{$\scriptstyle\sim$}}1400 t/y) and almost equal to the estimated reserves (24000 t). \cite{USGS2014} The required production for Se in 2030 ({\raise.17ex\hbox{$\scriptstyle\sim$}}14700 t/y) is more than twice the potential recoverable amount ({\raise.17ex\hbox{$\scriptstyle\sim$}}5500 t/y). The estimated Se reserves (120000 t) \cite{USGS2014} would be sufficient for around 8 years, at the required 2030 annual production levels.

The amount of Cd that is potentially recoverable from zinc ores can be estimated to be about 40500 t/y, based on the average Cd content of the zinc ore, sphalerite (ZnS), \cite{Bradshaw2013} and the annual zinc production of 13.5 million tons in 2012 \cite{USGS2013}. For Cd, the required production in 2030 for the Solar Generation 6 scenario \cite{EPIA2011} ({\raise.17ex\hbox{$\scriptstyle\sim$}}42300 t/y) is also above the potential production ({\raise.17ex\hbox{$\scriptstyle\sim$}}41000 t/y); however, the difference is proportionately less compared to In, Te, and Se. If the required level of annual Cd production is sustained, the estimated Cd reserves (500000 t) \cite{USGS2014} would be sufficient for around 12 years. 

Ga has the highest crustal abundance among all of the byproduct metals analyzed in this paper. Ga availability can be estimated to be 12500 t/y based on the average content of bauxite ores, \cite{Ayres2013} and the annual bauxite production in 2012, 250 million tons. \cite{USGS2013} The required Ga production in 2030 for the Solar Generation 6 scenario \cite{EPIA2011} ({\raise.17ex\hbox{$\scriptstyle\sim$}}2500 t/y) is almost an order of magnitude below the estimated maximum recoverable Ga based on today's bauxite production levels ({\raise.17ex\hbox{$\scriptstyle\sim$}}12500 t/y), and much lower than the estimated Ga reserves (400000 t). \cite{KirkOthmerGa}

Unlike the byproduct metals discussed above, Si is abundant: it comprises about 28\% of the Earth's crust as a constituent of various minerals. \cite{KirkOthmerSi} Although the U.S. Geological Survey does not report quantitative estimates of Si reserves, it states that the reserves are ample. \cite{USGS2014} If the annual Si production grows at its historical average annual growth rate of 2.2\% (as shown in figure \ref{fig:fig1}), then the annual Si production in 2030 will reach 11.5 million t/y in 2030. This is only 6\% lower than the annual Si production level required by the Solar Generation 6 scenario \cite{EPIA2011} in 2030, which is 12.2 million t/y.

%%%%%%%%%%%%%%%%%%%%%%%%%%%%%%%%%%%%%%%%%%%%%%%%%%%%%%%%%%%%%%
\section{Conclusion} 
%\input{conclusions}

% POINT: Summarize the paper shortly.
Continued rapid growth in PV deployment could require significant growth in the supply of some metals. In this paper, we estimate the growth rates needed in metals production to match PV deployment projections in 2030 for a range of future energy scenarios. We compare the required growth rates for six PV metals (In, Ga, Se, Te, Cd, and Si) with the historical growth rates observed for a large set of metals. We also compare the required production levels of the byproduct metals to their scalability potential based on metals reserves estimates.

% POINT: Emphasize the main result that In, Ga, Te, Se face high projected growth rates, while Si and Cd do not. 
The annual growth rates required for the byproduct metals (In, Ga, Te, and Se) production to satisfy the energy scenario-projected PV demand levels in 2030 are either unprecedented or fall on the higher end of the historical growth rates distribution. Growth projections for CdTe \{CIGS\} to supply 3\% \{10\%\} or greater electricity demand by 2030 would require unprecedented metals production growth rates for Te \{In\}. %, the required growth rates exceed the historical rates under the Solar Generation 6 scenario \cite{EPIA2011} scenario and medium material intensity. For greater PV adoption, Ga and Se growth rates would reach unprecedented growth rates as well. 
These estimates are for the medium material intensity case. The required metals growth rates will be even higher if material intensity remains at today's levels, the `high materials intensity' case. %(1) if the material intensity is not improved from current levels, which correspond to the high intensity case, or (2) under an energy scenario where CIGS or CdTe is assumed to provide a larger share of the electricity generation in 2030. Si and Cd, on the other hand, require growth rates that are on the lower end of the historical growth rates observed for all metals. 
In contrast, our results suggest that c-Si technology can provide up to 100\% of global electricity in 2030 without Si production exceeding the historical growth rates observed across a large set of metals. 

% POINT: Summarize the scalability potential of metals. 
The scalability potential of In, Te, and Se also fall short of the required production levels for these metals in 2030 based on estimated metals reserves. %This shows that it may not be possible to sufficiently expand the production of these metals by using their conventional sources even in the long run. 
Ga has a higher scalability potential based on its higher abundance in bauxite ores. The Cd supply does not appear to be constraining because of its higher abundance in ores and decreasing demand by non-PV uses due to its toxicity. Finally, Si supply restrictions do not appear to pose a binding scalability constraint, due to the abundance of this metal. 

% POINT: need for other PV technologies
This paper focuses on three main PV technologies that have been commercialized, CdTe, CIGS and c-Si. We find that at least one of these technologies, c-Si, is scalable based on the analysis of required metals production growth rates and Si availability. When the high processing costs of Si are considered, there is still room for improvement and possibly the introduction of non-Si based PV technologies, in order to reduce module costs. This study highlights, however, the importance from a scalability perspective of reducing the material intensity of other PV technologies (for example by using concentrators), or utilizing earth abundant materials. These general insights apply to a range of existing and future PV technologies.

% POINT: Limitation of the study: only primary production
% A source of uncertainty in this analysis is that we only focus on primary metal sources, that is mining the metals from the ores. We do not take into acccount the recycled metal obtained during the manufacturing processes or after the lifetime of the products. Systematic recycling of mine tailings, slag, and landfills can offer a secondary source of metals and increase the supply \cite{Hageluken2012}. This can reduce the growth needed especially for the primary sources of byproduct metals.

% POINT: Limitation of the study: non-PV uses growth rates
In this paper, the required metals growth rates reported rely on estimates of the future demand for non-PV end-uses, based on the range of observed historical growth rates of these metals. We note that if non-PV end-uses grow more rapidly or slowly than observed historically, the comparison of projected and historical growth rates to meet PV scenarios would change. %This is a source of uncertainty in our analysis. %Thus, we introduced a confidence interval for our estimations in order to account for the uncertainty about the future demand and the variability in the historical growth rates. 

The analysis of required growth rates in the context of historical growth rates provides a new perspective for assessing the raw material needs for future energy deployment scenarios. This approach can also be useful for analyzing the materials requirements of other technologies, to assess their scalability and inform technology development and research investment. We note that while the availability of raw materials is a necessary condition for scaling up technology production, other factors including production energy requirements \cite{Darling2013,Yue2014,Yue2014corr} and the regional distribution of resources \cite{Goetze2014}, should also be considered in an analysis of sufficient conditions for scalability.

%%%%%%%%%%%%%%%%%%%%%%%%%%%%%%%%%%%%%%%%%%%%%%%%%%%%%%%%%%%%%%
\section*{Acknowledgment}
We thank the DOE for supporting this research under grant DE-EE0006131.

%%%%%%%%%%%%%%%%%%%%%%%%%%%%%%%%%%%%%%%%%%%%%%%%%%%%%%%%%%%%%%


\balance


\footnotesize{
%\bibliography{references} %your .bib file
%\bibliographystyle{rsc} %the RSC's .bst file
%}

\providecommand*{\mcitethebibliography}{\thebibliography}
\csname @ifundefined\endcsname{endmcitethebibliography}
{\let\endmcitethebibliography\endthebibliography}{}

}

\end{document}